\documentclass[prd,aps,twocolumn,floats,floatfix,nofootinbib] {revtex4-1}
\usepackage{lipsum}
\usepackage{graphicx}
\usepackage{dcolumn}
\usepackage{bm}
\usepackage{graphics}
\usepackage{slashed}
\usepackage{amssymb}
\usepackage{natbib}
\usepackage{amsmath}
\usepackage{url}
\usepackage{multirow}
\usepackage[usenames,dvipsnames,svgnames,table]{xcolor}
\usepackage{color}
\usepackage{xcolor}
\usepackage{tabularx}
\usepackage{hyperref}

\def\beq{\begin{equation}}
\def\eeq{\end{equation}}
\newcommand{\bea}{\begin{eqnarray}}
\newcommand{\eea}{\end{eqnarray}}

\newcommand{\tp}{\tilde{p}}
\newcommand{\trho}{\tilde{\rho}}
\newcommand{\teps}{\tilde{\epsilon}}
\newcommand{\teinf}{\tilde{\epsilon}_{\infty}}

\begin{document}

\title{Dynamical Chameleon Neutron Stars: stability, radial oscillations and scalar radiation in spherical symmetry}

  \author{Alexandru Dima}
 \affiliation{SISSA, Via Bonomea 265, 34136 Trieste, Italy and INFN Sezione di Trieste}
 \affiliation{IFPU - Institute for Fundamental Physics of the Universe, Via Beirut 2, 34014 Trieste, Italy}
 \author{Miguel Bezares}
 \affiliation{SISSA, Via Bonomea 265, 34136 Trieste, Italy and INFN Sezione di Trieste}
 \affiliation{IFPU - Institute for Fundamental Physics of the Universe, Via Beirut 2, 34014 Trieste, Italy}
  \author{Enrico Barausse}
 \affiliation{SISSA, Via Bonomea 265, 34136 Trieste, Italy and INFN Sezione di Trieste}
 \affiliation{IFPU - Institute for Fundamental Physics of the Universe, Via Beirut 2, 34014 Trieste, Italy}

\begin{abstract}
Scalar-tensor theories whose phenomenology differs significantly from general relativity
on large (e.g. cosmological) scales do not typically pass local experimental tests (e.g.
in the solar system) unless they present a suitable ``screening mechanism''.
An example is provided by chameleon screening, whereby the local general relativistic behavior is
recovered in high density environments, at least in weak-field and quasi-static configurations.
Here, we test the validity of chameleon screening in strong-field and highly relativistic/dynamical conditions,
by performing fully non-linear simulations of neutron stars subjected to
initial perturbations that cause them to oscillate or even collapse to a black hole.
We confirm that screened chameleon stars are stable to sufficiently small radial
oscillations, but that the frequency spectrum of the latter shows deviations from the general relativistic predictions.
We also calculate the scalar fluxes produced during collapse to a black hole, and comment on their detectability with future
gravitational-wave interferometers.
\end{abstract}

\maketitle

\section{Introduction}

Astrophysical compact objects, such as neutron stars (NSs) and black holes (BHs), offer an exceptional laboratory to test gravity in the strong field regime and constrain extensions of General Relativity (GR)~\cite{Damour:1996ke,Will:2014kxa,Berti:2015itd,Barausse:2016eii,Berti:2016lat,TheLIGOScientific:2016src,PhysRevLett.123.011102,will_2018,Berti:2018cxi,Berti:2018vdi,LIGOScientific:2019fpa,Barausse:2020rsu,Abbott:2020jks,Volkel:2020xlc}.
The most studied extensions of GR are scalar-tensor (ST) theories of gravity~\cite{JordanPascual1952SuWG,Fierz:1956zz,Jordan:1959eg,Brans:1961sx,Damour:1992we,Fujii:2003pa}, which introduce one (or more~\cite{Damour:1992we,Horbatsch:2015bua}) scalar field(s) that mediate the gravitational interaction (together with the metric tensor).
These theories may have applications in cosmology (at both early and late times), and scalar fields may even play the role of dark matter~\cite{CLIFTON20121,Arvanitaki:2010sy,Arvanitaki:2015iga,Brito:2017wnc,Brito:2017zvb,PhysRevD.97.075020}, although
agreement with both local and cosmological scales is not always easy to ensure.

Because of \textit{no-hair} theorems~(see~\cite{Bekenstein:1996pn,Herdeiro:2015waa,Sotiriou:2015pka} for reviews), a broad class of ST theories do not leave any characteristic imprint in the physics of vacuum solutions (with
the exception of theories allowing for BH scalarization~\cite{Sotiriou:2013qea,Sotiriou:2014pfa,Silva:2017uqg,Dima:2020yac,Herdeiro:2020wei}).
However, although no-hair theorems are known to exist also for stars in certain classes of ST theories~\cite{Barausse:2015wia,Yagi:2015oca,Lehebel:2017fag}, non-vacuum spacetimes are generally regarded as  more promising testing grounds for extensions of GR, because
deviations from GR are enhanced by the modified coupling between matter and gravity. In particular, ST theories typically introduce a coupling of the scalar gravitational field to the trace of the stress-energy tensor, which can produce non-perturbative effects such as \textit{ scalarization}~\cite{PhysRevLett.70.2220,Damour:1996ke,Barausse:2012da,Palenzuela:2013hsa,PhysRevD.89.084005,Sennett:2017lcx}.
In fact, even when these theories satisfy the constraints coming from solar system tests~\cite{2003Natur.425..374B,Murphy:2012rea,Will:2014kxa}, they can predict measurable deviations from GR in the structure, dynamics and radiative emissions of NSs~\cite{Freire:2012mg,PhysRevD.91.064024,PhysRevD.93.044009,PhysRevD.90.124091,PhysRevD.93.124035,PhysRevD.89.084005,Soldateschi:2020zxb}.

Particularly interesting is the existence of classes of ST theories that are endowed with \textit{screening mechanisms} devised to hide non-GR effects on astrophysical (local) scales, while leaving room for modifications on cosmological ones~\cite{Joyce:2014kja}.  Known examples of these mechanisms include: \textit{kinetic} screening (\textit{k-mouflage}~\cite{Babichev:2009ee,2011PhRvD..84f1502B,Brax:2012jr,Burrage:2014uwa}); \textit{Vainshtein} screening~\cite{VAINSHTEIN1972393,Deffayet:2001uk,Babichev:2009us,Babichev:2010jd}; screening based on an \textit{environmentally weak coupling of the scalar field to  matter} (symmetron~\cite{Pietroni:2005pv,Olive:2007aj,Hinterbichler:2010es} or dilaton models~\cite{Damour:1994zq,PhysRevD.83.104026});  or an \textit{environmentally large mass} of the scalar field, as in chameleon screening~\cite{Khoury:2003rn,PhysRevLett.93.171104}.

Chameleon screening is indeed realized
by endowing the scalar degree of freedom with an effective mass that depends on the ambient matter density:
in high-density environments (e.g. compact objects, our solar system or even galaxies and clusters) small perturbations are suppressed by the large inertia
of the field, while on larger cosmological scales  lower densities allow for \textit{quintessence}-like effects, arising from to the non-trivial self-interaction potential~\cite{PhysRevLett.93.171104}.
Moreover, the scalar charge of compact objects receives contributions only from a small volume located close to the surface: this \textit{thin-shell effect} effectively suppresses the scalar force~\cite{Khoury:2003rn}.

Screening mechanisms  generally make modifications of gravity elusive and hard to constrain with astrophysical observations. Nonetheless, their efficacy at screening compact stars is typically tested in the static non-relativistic limit, and little work has been done outside these simplifying approximations (e.g. see~\cite{terHaar:2020xxb,Bezares:2021yek} for the dynamics of k-mouflage). This is also the case for chameleon screening, the robustness of which has only been tested so far in the dynamical Newtonian limit~\cite{Nakamura:2020ihr}, or in the relativistic but static regime~\cite{PhysRevD.81.124051,PhysRevD.95.083514,deAguiar:2020urb} (see also ~\cite{Sagunski:2017nzb,Lagos_2020} for other relevant work on chameleon screening).
In this regard, one potential loophole in chameleon screening could be opened by a tachyonic instability developing inside relativistic compact stars. This instability
arises in ST theories without screening~\cite{PhysRevD.91.064024}, where it leads either to scalarization or alternatively to gravitational collapse~\cite{PhysRevD.93.124035}.
Past work~\cite{PhysRevD.81.124051,PhysRevD.95.083514} reported instabilities of the chameleon field inside neutron stars with a pressure-dominated core. These instabilities were interpreted as due to the chameleon effective potential not having a well defined minimum for the scalar field to relax to, as a consequence of the trace of the matter stress-energy tensor changing sign in the highly relativistic interior of the stars. Recently, however, Ref.~\cite{deAguiar:2020urb} has studied static NS solutions coupled to chameleon scalar fields and, in contrast to previous work, found no sign of such instabilities. Instead, they observed that NSs with pressure-dominated cores typically present a partial \textit{descreening} in their interior and are linearly stable. Many realistic candidates
for the equation of state (EoS) of nuclear matter predict pressure-dominated cores at sufficiently high densities, while
agreeing with current experimental constraints~\cite{Podkowka:2018gib}. One may therefore place
bounds on theories with chameleon screening from observations of the most massive NSs.

As our first main contribution, in this work
we will confirm and generalize the conclusions obtained in Ref.~\cite{deAguiar:2020urb}, which are in principle valid only at the level of linear perturbations around static solutions. We will do that by demonstrating numerically the long-term nonlinear stability of NS solutions coupled to a chameleon scalar field, which we will henceforth refer to as chameleon NSs (CNSs). To our knowledge, these are the first dynamical simulations of the chameleon screening mechanism, thanks to which we confirm that the partial descreening inside pressure-dominated cores leads to stable CNSs that deviate strongly from GR.

As is well know, in GR radial oscillations of relativistic stars do not source gravitational wave (GW) emissions (although in principle they can couple to non-radial modes~\cite{Passamonti:2004je,Passamonti:2005cz,PhysRevD.75.084038} and potentially be observable during the post-merger phase~\cite{10.1111/j.1365-2966.2011.19493.x,PhysRevD.91.064001,Vretinaris:2019spn}). For this reason, they are typically studied only for assessing the stability of NS solutions~\cite{PhysRevLett.12.114,1964ApJ...140..417C,1966ApJ...145..514M,1983ApJS...53...93G,Kokkotas:RadOsc}.
However, in ST theories a new family of modes typically appears in association with the additional degree of freedom~\cite{Mendes:2018qwo,Blazquez-Salcedo:2020ibb}. These radial modes can source the emission of (scalar) GWs~\cite{Sotani:2014tua} (for instance, during collapse~\cite{Gerosa:2016fri,Sperhake:2017itk,Rosca-Mead:2020ehn,Bezares:2021yek}).
In this work, we study the spectrum of radially perturbed CNSs,  characterizing the deviations from GR induced by the chameleon field. In addition, we compute the scalar flux radiated by CNSs when oscillating or collapsing to a BH, focusing on the comparison between screened and descreened stars and on the observability with current and future GW detectors.

This paper is organized as follows. In Sec.~\ref{two} we briefly review chameleon gravity and its screening mechanism. We also discuss the current constraints and the relevance of these theories for cosmological applications. In Sec.~\ref{three}, we discuss the initial data that are used in our simulations and the numerical method employed to produce them.
The evolution formalism is presented in Sec.~\ref{four}, where we also discuss the stability of CNSs. In Sec.~\ref{five} we discuss characteristic radial oscillations of CNSs and in Sec.~\ref{six} we characterize the monopole emission of oscillating and collapsing CNSs. Finally, in Sec.~\ref{conclusion} we discuss our conclusions and the future prospects to test chameleon screening with NSs.
Throughout this paper, we use natural units where $\hbar=c=1$.

\section{Theoretical framework}\label{two}
\subsection{Screened modified gravity action}

ST theories with environmentally dependent screening,
such as symmetron, dilaton or chameleon screening (including certain $f(R)$ models),
are described by the following action~\cite{PhysRevD.86.044015}:
\begin{align}\label{eq:action}
S=&\int d^4x\sqrt{-g}\left[\frac{M_{\textrm{pl}}^2}{2}R-\frac{1}{2}g^{\mu\nu}\nabla_{\mu}\phi\nabla_{\nu}\phi-V(\phi)\right]\\
&+S_m[A(\phi)^2g_{\mu\nu};\psi_m]\nonumber\,,
\end{align}
where $g$ and $R$ are the determinant and Ricci scalar of the \textit{Einstein frame} metric
$g_{\mu\nu}$, and $M_{\textrm{pl}}=1/\sqrt{8\pi G}$ is the (reduced) Planck mass. The scalar field $\phi$ has a
self-interaction potential $V(\phi)$, and is coupled to matter (collectively
represented by the field $\psi_m$) through the conformal coupling $A(\phi)$.
Because of this coupling, matter does not follow geodesics of $g_{\mu\nu}$,
but ones of the \textit{Jordan frame} metric~\cite{PhysRevD.1.3209}
\begin{equation}\label{eq:jordan}
\tilde{g}_{\mu\nu}\equiv A(\phi)^2g_{\mu\nu}\,.
\end{equation}

Therefore, in this frame one can define a stress-energy tensor,
\begin{equation}\label{eq:set}
\tilde{T}_{\mu\nu}^{m}\equiv -\frac{2}{\sqrt{-\tilde{g}}}\left(\frac{\delta S_m}{\delta \tilde{g}^{\mu\nu}}\right)\,,
\end{equation}
and a baryon mass current, $\tilde{J}^{\mu}$, that are covariantly conserved
\begin{align}
\tilde{\nabla}_{\mu}\tilde{J}^{\mu}&=0\,,\label{eq:consJJ}\\
\tilde{\nabla}_{\mu}\tilde{T}^{\mu\nu}_m&=0\label{eq:consTJ}\,,
\end{align}
where $\tilde{\nabla}$ indicates the covariant derivative compatible with the Jordan frame metric~\eqref{eq:jordan}. In this work, the matter content of the spacetime is modeled as a perfect fluid in the Jordan frame, with stress-energy tensor
\begin{equation}\label{eq:setm}
\tilde{T}^{\mu\nu}_{m}\equiv (\tilde{\epsilon}+\tilde{p})\tilde{u}^{\mu}\tilde{u}^{\nu}+\tilde{p}g^{\mu\nu}\,.
\end{equation}
The Jordan-frame fluid variables in this equation (total energy density, $\teps$, and isotropic pressure, $\tilde{p}$) are defined as measured by an observer comoving with the fluid elements with four-velocity $\tilde{u}^{\mu}$.

By defining the Einstein-frame stress-energy tensor as $T^{\mu\nu}_{m}\equiv-2/\sqrt{-g}\left(\delta S_m/\delta g^{\mu\nu}\right)$ and comparing the latter with~\eqref{eq:set}, one obtains the relation $T^{m}_{\mu\nu}=A(\phi)^{2}\tilde{T}^{m}_{\mu\nu}$. From this conformal transformation, and from $u^{\mu}=A(\phi)\tilde{u}^{\mu}$ (obtained from the normalization $g_{\mu\nu}u^{\mu}u^{\nu}=-1$) one can obtain the correspondence between fluid variables in the two frames, $\epsilon=A(\phi)^4\tilde{\epsilon}$ and $p=A(\phi)^4\tilde{p}$.
The conserved Jordan-frame baryon mass current, $\tilde{J}^{\mu}\equiv\tilde{\rho}\tilde{u}^{\mu}$, where $\tilde{\rho}$ is the rest-mass density, is related to the corresponding Einstein-frame quantity by $J^{\mu}=A(\phi)^{5}\tilde{J}^{\mu}$~\cite{Palenzuela:2013hsa}.
Note that in the Einstein frame covariant conservation of the stress-energy tensor and baryon mass current is lost and Eqs.~\eqref{eq:consJJ} and ~\eqref{eq:consTJ} are replaced by:
\begin{align}
\nabla_{\mu}J^{\mu}=\frac{d\ln A(\phi)}{d\phi}J^{\mu}\nabla_{\mu}\phi\,,\label{eq:consJ}\\
\nabla_{\mu}T^{\mu\nu}_m=\frac{d\ln A(\phi)}{d\phi}T_m\nabla^{\nu}\phi\,,\label{eq:consT}
\end{align}
where $T^m=g^{\mu\nu}T^m_{\mu\nu}$ is the trace of the stress-energy tensor.

Variation of the action~\eqref{eq:action} with respect to the Einstein metric gives the modified Einstein field equations
\begin{equation}
G_{\mu\nu}=8\pi G\left(T_{\mu\nu}^{\phi}+T_{\mu\nu}^{m}\right)\label{eq:einstein}~,
\end{equation}
which are sourced by the stress-energy tensor of the scalar field
\begin{equation}\label{eq:setphi}
T_{\mu\nu}^{\phi}\equiv \nabla_{\mu}\phi\nabla_{\nu}\phi-g_{\mu\nu}\left(\frac{1}{2}\nabla_{\sigma}\phi\nabla^{\sigma}\phi+V(\phi)\right)\,.
\end{equation}
The scalar field equation is obtained by variation of~\eqref{eq:action} with respect to $\phi$:
\begin{equation}
\Box \phi =\frac{dV(\phi)}{d\phi}-\frac{d\ln A(\phi)}{d\phi}T^m~,\label{eq:kg}
\end{equation}
which is a generalized wave equation on curved spacetime with $\Box\equiv g^{\mu\nu}\nabla_{\mu}\nabla_{\nu}$, sourced by the scalar self-interaction and by the coupling to the Einstein-frame trace of the stress-energy tensor.

Specifying $V(\phi)$ and $A(\phi)$ one specializes to a particular model of chameleon gravity. In this work, we will focus on the classic chameleon models that feature an inverse power-law self-interaction potential in combination with an exponential conformal coupling to matter, i.e.:
\begin{equation}\label{eq:chameleon}
V(\phi)=\frac{\Lambda^{n+4}}{\phi^n}~, \hspace{0.5cm} A(\phi)=\text{exp}\left(\alpha_0\phi \right)\,,
\end{equation}
where $\Lambda$ is the chameleon energy scale and $\alpha_0$ is the dimensionful conformal coupling. Plugging~\eqref{eq:chameleon} into~\eqref{eq:kg}, one can see that the chameleon scalar field obeys an effective potential
\begin{equation}\label{eq:effpot}
V_{\text{eff}}(\phi)\equiv \frac{\Lambda^{n+4}}{\phi^n}- \frac{1}{4}e^{4\alpha_0\phi}\tilde{T}_m\,.
\end{equation}
In this paper, we consider only the simplest chameleon model $n=1$.
The scalar configuration that minimizes the potential~\eqref{eq:effpot}, $\bar{\phi}$, can be found by requiring $dV_{\textrm{eff}}/d\phi|_{\bar{\phi}}=0$ or, equivalently, by solving the trascendental equation $\phi^2e^{4\alpha_0\phi}+\Lambda^5/(\alpha_0\tilde{T}_m)=0$, which in the limit $\phi\ll M_{\textrm{pl}}$ yields
\begin{equation}\label{eq:phimin}
\bar{\phi}\simeq\frac{\Lambda^{5/2}}{\sqrt{-\alpha_0\tilde{T}_m}}\,.
\end{equation}
From the effective potential~\eqref{eq:effpot} one can determine the chameleon effective mass,
\begin{equation}\label{eq:mass}
m_{\textrm{eff}}^2\equiv\frac{d^2V_{\textrm{eff}}}{d\phi^2}=\frac{2\Lambda^5}{\phi^3}-4\alpha_0^2e^{4\alpha_0\phi}\tilde{T}_m\,.
\end{equation}

\subsection{Chameleon screening}~\label{twoB}
The field configuration that minimizes the effective potential~\eqref{eq:effpot} strongly depends on the ambient matter distribution: in denser regions the chameleon will settle to lower field values and scalar perturbations around the minimum will feature a larger effective mass~\eqref{eq:mass}. As a result, the chameleon fifth force
will be short-range in high-density environments (i.e. stars, clusters or galaxies), while being effectively long-range on cosmological scales. In addition, a \textit{thin-shell effect} will further suppress the fifth force around compact objects (e.g. NSs~\cite{PhysRevLett.93.171104}).

As an illustrative example, let us consider a non-relativistic, static and spherical star of mass $M$ and radius $R$, surrounded by a medium (e.g. the interstellar medium, or even the cosmological background) with lower density, $\teinf$. Inside the star and far from it, the chameleon  will settle to different field values. The large effective mass, corresponding to the high density in the interior, will suppress exponentially the scalar perturbations and keep the chameleon field small up to a \textit{screening radius}, $r_s$. The latter can be defined as the distance from the center at which the field starts rolling towards the ``exterior'' minimum. Inside the screening radius, the gradient of the scalar field is negligible and the fifth force (proportional to the gradient) reactivates only outside of it, $r\gtrsim r_s$. One can show that sufficiently far away from the star, at $r\gg R_{star}$, the scalar field solution is
\begin{equation}\label{eq:phinf}
\phi\simeq \phi_{\infty}-\left(\frac{Q}{4\pi M_{\textrm{pl}}}\right)\frac{M e^{-m_{\infty}(r-R_{star})}}{r}\,,
\end{equation}
with $Q$ being the (dimensionless) effective scalar charge of the object and $m_{\infty}$ the chameleon effective mass~\eqref{eq:mass} at large distances. From \eqref{eq:phinf} one can notice that the chameleon mass term introduces an exponential suppression of the ``Yukawa'' type.

In the non-relativistic Newtonian limit the charge reads $QM\simeq\alpha_0M_{\textrm{pl}}(M-M(r_s)\big)$, where $M(r_s)$ is the gravitational mass contained inside the screening radius~\cite{Khoury:2003rn,Hui:2009kc}.
When the star is efficiently screened, i.e. $r_s\sim R$,
the scalar charge is only sourced by a ``thin shell'' of matter between $r_s$ and $R$, and the fifth force is additionally suppressed by the factor $Q\ll 1$~\cite{PhysRevLett.93.171104,Khoury:2003rn,Sakstein:2016oel,Sakstein:2018fwz}.
As long as $T<0$ (which in the non-relativistic limit is automatically satisfied), the chameleon effective potential~\eqref{eq:effpot} has
a minimum in the stellar interior, and this thin-shell effect is present. However,
in the pressure-dominated core of very dense NSs, $T$ can change sign, leading to
a partial breakdown of chameleon screening~\cite{deAguiar:2020urb}. In this paper,
we will explore the dynamics of this breakdown, or \textit{descreeeing}.

\subsection{Constraints}
Although it has been demonstrated that chameleon scalar fields cannot give rise to self-acceleration~\cite{Wang:2012kj}, they could still be relevant for cosmological applications in combination with a cosmological constant, as both could have a common origin at high energies~\cite{PhysRevD.70.123518}. Indeed, the low-energy effective theories derived from string theory are generically populated with light scalar fields and the chameleon screening might be a viable mechanism
to hide their presence in experiments. In this perspective, relatively recent work has found that chameleon models are compatible with the swampland program, provided that a lower bound on the conformal coupling is satisfied~\cite{Brax:2019rwf}.

However, while not completely ruled out yet, classic chameleon models are constrained by a variety of observations (see~\cite{Burrage:2017qrf,Sakstein:2018fwz,Baker:2019gxo} for reviews). The viable region of the parameter space of the most studied chameleon model (i.e.~\eqref{eq:effpot} with $n=1$)
is $\alpha_0M_{\textrm{pl}} \lesssim O(10^{2})$ for energy scales $\Lambda\lesssim\Lambda_{\rm DE}$~\cite{Sakstein:2018fwz}, where $\Lambda_{\rm DE}=2.4$ meV is the Dark Energy scale.
Further constraints may come from the scales of galaxies/galaxy clusters, although they have not been worked out in detail~\cite{Desmond:2020gzn},
and from short-range experiments~\cite{Pernot-Borras:2021edr}.

\section{Initial Data}\label{three}
In this section, we derive static and spherically symmetric solutions for CNSs, by generalizing the Tolman-Oppenheimer-Volkoff (TOV) equations
to the chameleon case and solving them numerically. We also discuss the EoS of nuclear matter and the boundary conditions used, and present results for the mass-radius relation of CNSs.
\subsection{Chameleon-TOV equations}
To obtain the  modified TOV equations, we adopt the following spherically symmetric ansatz (in polar coordinates):
\begin{equation}\label{eq:ansatz}
ds^2=-e^{2\nu(r)}dt^2+e^{2\lambda(r)}dr^2+r^2(d\theta^2+\sin^2\theta d\varphi^2)\,.
\end{equation}
By inserting the ansatz~\eqref{eq:ansatz} in the chameleon field equations~\eqref{eq:einstein} and \eqref{eq:kg}, one obtains
\begin{eqnarray}
\frac{d\nu}{dr}&=&\left(\frac{e^{2\lambda}-1}{2r}\right)+ \nonumber\\
&&\quad+4\pi Gr e^{2\lambda}\left(  e^{4\alpha_0\phi}\tp + \frac{\Lambda^5}{\phi}-\frac{e^{-2\lambda}}{2}\sigma^2\right)\label{eq:nu}\,,\\
\frac{d\lambda}{dr}&=&\left(\frac{1-e^{2\lambda}}{2r}\right)+\nonumber\\
&&\quad+4\pi Gr e^{2\lambda}\left( e^{4\alpha_0\phi}\teps- \frac{\Lambda^5}{\phi}-\frac{e^{-2\lambda}}{2}\sigma^2\right)\label{eq:lambda}\,,\\
\frac{d\sigma}{dr}&=&\left( \frac{d\lambda}{dr}-\frac{d\nu}{dr}-\frac{2}{r}\right)\sigma+\nonumber\\
&&\quad-e^{2\lambda}\left(\alpha_0e^{4\alpha_0\phi}(3\tp-\teps)+\frac{\Lambda^5}{\phi^2}\right)\label{eq:sigma}\,,\\
\frac{d\phi}{dr}&=&\sigma~,\label{eq:phi}\\
\frac{d\tp}{dr}&=&-(\tp+\teps)\left(\frac{d\nu}{dr}+\alpha_0\sigma\right)\label{eq:pres}\,.
\end{eqnarray}
The differences from the TOV equations in GR depend on the conformal coupling $\alpha_0$ and the chameleon energy scale $\Lambda$, both introduced in Eq.~\eqref{eq:chameleon}.
This system of equations can be solved numerically by using suitable boundary conditions and choosing an adequate EoS for nuclear matter, as we explain in detail in the next subsection.

\subsection{Equation of state and boundary conditions}

To close the system of equations~\eqref{eq:nu},~\eqref{eq:lambda},~\eqref{eq:sigma},~\eqref{eq:phi} and~\eqref{eq:pres},  a relation between the fluid variables  must be provided. We choose to describe the stellar interior with a polytropic EoS
\begin{equation}\label{eq:polyEoS}
\tp(\trho)\equiv K\trho^{\Gamma}\,, \hspace{1cm}  \teps(\tp)=\frac{\tp}{\Gamma-1}+\left(\frac{\tp}{K}\right)^{1/\Gamma}~\,,
\end{equation}
where $K$ is the polytropic constant and $\Gamma$ is the (constant) adiabatic index.
This EoS, while approximate, allows for reproducing
the relativistic effects found in pressure-dominated NS cores (e.g., see~\cite{PhysRevD.93.124035} for an application to scalarized NSs) for appropriately stiff polytropic coefficients~\cite{Podkowka:2018gib}.
In this paper, we generally set $\Gamma=3$,
which approximates the polytropic exponent of more realistic EoSs \cite{PhysRevD.79.124032},
and $K\simeq6.9\times10^{4} G^6M_{\odot}^4c^{-10}$.
In GR, this EoS yields a  maximum mass
$M_{\max}\simeq2.03 M_{\odot}$, consistent with current bounds~\cite{Cromartie:2019kug}.
As we will see, this stiff EoS yields static stars with a partially descreened interior.
We will also use a different polytropic EoS with $\Gamma=2$ and $K=123~G^3M_{\odot}^2c^{-4}$ to obtain CNSs with similar baryon mass but with a completely screened interior, for comparison (see section~\ref{five}).

Outside the star, $r\geq R_{star}$, we
assume a homogeneous atmosphere, $\teps=\teinf=\textrm{const.}$, with ``cosmological'' EoS, $\tp=-\teinf$, corresponding to a cosmological constant.
Chameleon models with a runaway potential such as that of Eq.~\eqref{eq:chameleon} do not admit a constant scalar field solution in pure vacuum, and for this reason a homogeneous atmosphere is required to have a well-behaved exterior solution. In fact, it is easy to see that with this cosmological atmosphere
the field equations allow for the asymptotic solution
$\teps= \teinf$, $\tp=-\teinf$, $\phi= \phi_{\infty}$ at  $r \gg R_{star}$, with $R_{star}$ being the radius of the star.
Once fixed the atmosphere density, $\teinf$, the asymptotic chameleon configuration, $\phi_{\infty}$, is determined by~\eqref{eq:phimin}, where $\tilde{T}_m=-4\teinf$.
Consistently, the metric is then given (asymptotically) by the Schwarzschild-de Sitter solution
\begin{equation}\label{eq:Schw-dS}
ds^{2} =-f(r)dt^2+f(r)^{-1}dr^2+r^2(d\theta^2+\sin^2\theta d\varphi^2)~,
\end{equation}
where $f(r)\equiv1-2GM/r-b\,r^2/3$, with $M$ the gravitational mass and $b=8\pi G V_{\text{eff}}(\teps_{\infty})|_{\phi_{\infty}}$. Instead, the (Jordan-frame) baryon mass of the star is defined as
\begin{equation}\label{eq:mbar}
M_{bar}\equiv\int d^3x\sqrt{-\tilde{g}}\tilde{J}^0\,.
\end{equation}
Other prescriptions for the atmosphere are possible, for instance in terms of a non-relativistic homogeneous dust distribution modeling the interstellar medium, but the advantage of choosing a cosmological EoS is
that it yields a simple exterior solution~\cite{deAguiar:2020urb}.

\subsection{Static Chameleon Neutron Stars}

Static and spherically symmetric CNS solutions are obtained numerically by integrating outwards the modified TOV equations starting from the center of the star, where we impose regular boundary conditions. To implement the Schwarzchild-de Sitter boundary conditions [Eq.~\eqref{eq:Schw-dS}] far away from the star, we
use a direct shooting method.

We consider atmosphere densities of order $\teinf\sim10^{-6}$--$10^{-4}\rho_{\text{nuc}}c^2$, where $\rho_{\text{nuc}}=1.7\times 10^{14}$ g/cm$^{3}$ is a typical nuclear density. Notice that our direct shooting method cannot handle more realistic atmospheres like a background cosmological density, $\rho_c\simeq 1.0\times 10^{-23}$ g/cm$^{3}$, or the density of the interstellar medium (in a giant molecular cloud~\cite{gmc}), $\rho_{GMC}\simeq1.7\times 10^{-20}$ g/cm$^{3}$. For the chameleon action parameters we set $\alpha_0M_{\textrm{pl}}= 1$
and $\Lambda\simeq 73$--$175$ GeV. These chameleon energy scales
are inconsistent  with current bounds~\cite{Sakstein:2018fwz}, but lower (and viable) values of $\Lambda$ are again impossible to explore with our shooting method. This is because to solve for CNSs we have to utilize code units adapted to the problem, where $G=c=M_{\odot}=1$. Reinstating all $\hbar$, $c$ and $G$ factors one obtains $\Lambda^5=(8\pi)^{-3/2}(M_{\textrm{pl}}/\hbar c)^2(GM_{\odot}/c^2)^2(\bar{\Lambda}/M_{\textrm{pl}})^5\approx 2.6\times 10^{72}~(\bar{\Lambda}/M_{\textrm{pl}})^5$, and realistic values of $\Lambda$ therefore become tiny and hard to handle numerically.
This is a problem commonly encountered when simulating compact stars in theories with screening (see e.g.~\cite{Brax:2012jr,deAguiar:2020urb,Bezares:2021yek}),
and it stems from the separation between the cosmological scale $\Lambda$ and that of NSs. In Sec.~\ref{six}, however,
we will extrapolate our results to more realistic values of both $\Lambda$ and $\teinf$ by using semi-analytic arguments.

\begin{figure}[th!]
\centering
\includegraphics[width=0.5\textwidth]{ 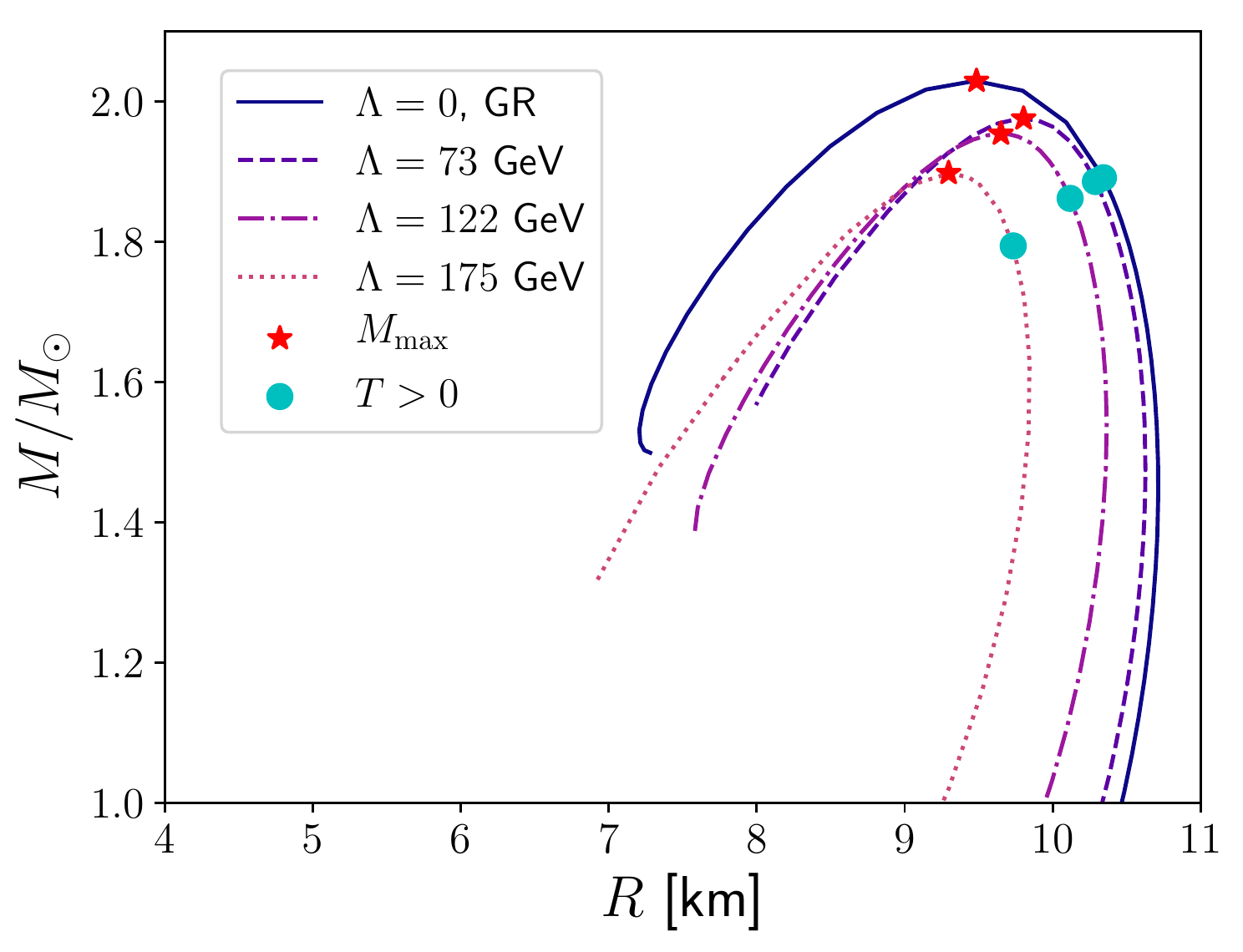}
\caption{{\em Mass-radius plots for varying chameleon energy scales.}
Dotted, dashed-dotted and dashed lines correspond to CNSs with $\alpha_0M_{\text{pl}}=1$ : darker lines correspond to lower chameleon energy scales. The darkest solid line corresponds to NS solutions in GR
($\alpha_0=0$ and $\Lambda=0$). Red star tokens indicate the solutions with maximum mass. Cyan round tokens, instead, indicate the lightest star featuring a pressure-dominated ($T>0$) core: stable CNSs with a partially descreened core are those between the star and round tokens.}
\label{fig:MassRadius}
\end{figure}

\begin{figure}[th!]
\centering
\includegraphics[width=0.5\textwidth]{ 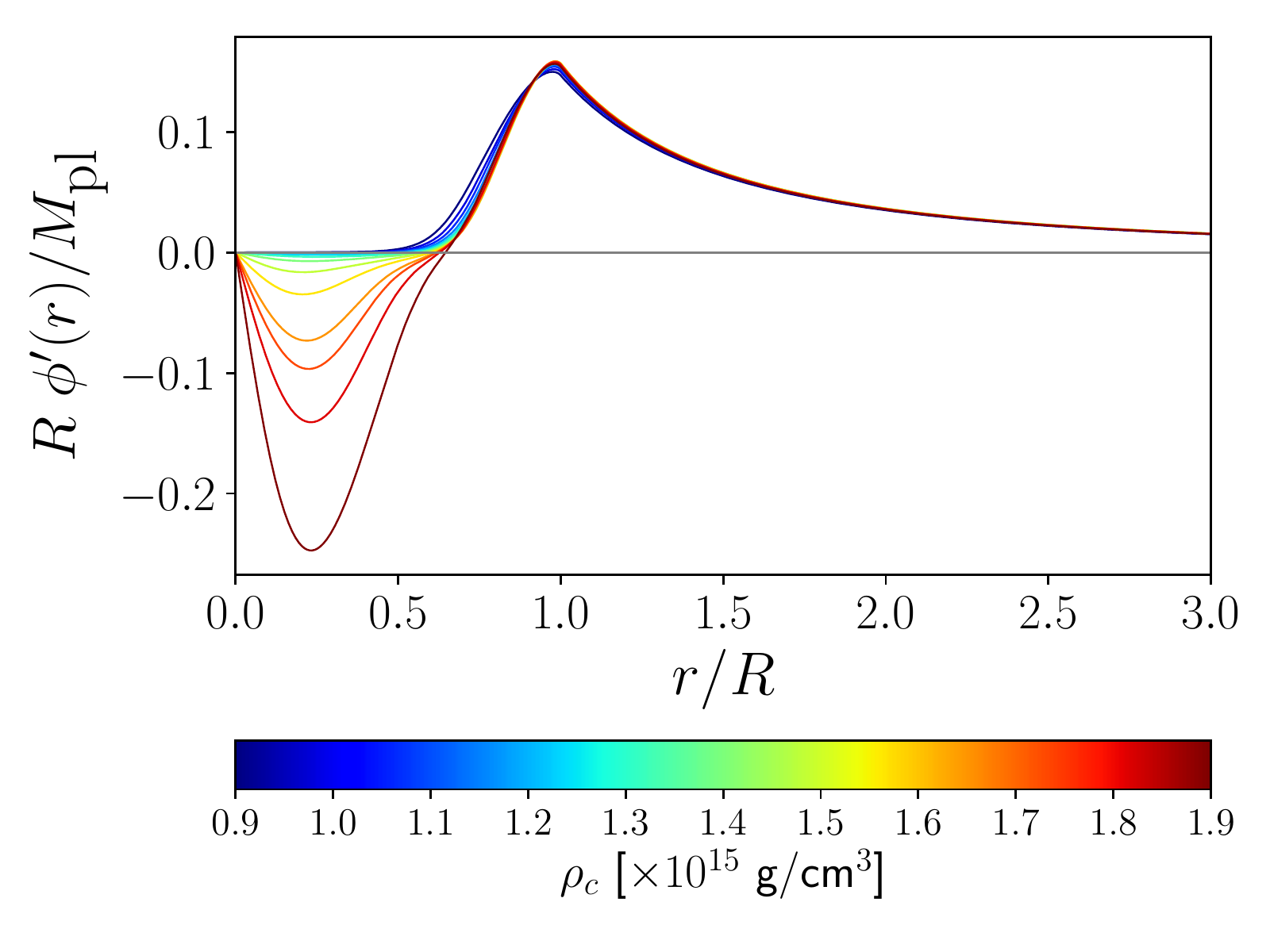}
\caption{ {\em{Partial descreening of CNSs.}} Gradient of the chameleon field around CNS solutions with varying central densities: less dense stars (blue end of the color scale) are screened in their interior, as can be seen from the suppression of the scalar field gradient; stellar solutions with higher central densities (red end of the color scale)
feature a scalar field gradient (proportional to the chameleon-propagated fifth force) reactivated in their interior as a result of relativistic effects, i.e. the pressure-dominated cores.}
\label{fig:descreen}
\end{figure}

Mass-radius curves for different values of $\Lambda$ are shown in Fig.~\ref{fig:MassRadius}, where
we also show the GR case ($\Lambda=0$).
These curves are comprised of stable and unstable stars,
which lie, respectively, on the right of the maximum mass configuration (red star tokens) and on its left. Additionally,
solutions between the red star token and the cyan round token
have $T>0$ (pressure dominated core).
As mentioned previously, in chameleon gravity, a pressure-dominated core can produce a partial \textit{descreening}.
As can be observed from Fig.~\ref{fig:descreen}, that consists of a re-activation
of the scalar gradient (and thus of the fifth force) in the stellar interior, where it would normally be suppressed by screening.
For fixed mass, screened CNSs are typically smaller in size and more compact than NSs in GR. However, in the limit $\Lambda\rightarrow 0$ screened solutions tend smoothly to GR configurations. Descreened solutions, instead,
feature strong deviations, as can be observed from the fact that the maximum mass is typically lower than in GR and in the limit $\Lambda\rightarrow 0$ the most massive GR configuration is not recovered smoothly. Moreover, the branch of unstable solutions shows the strongest structural deviations from GR, even for smaller chameleon energy scales $\Lambda$.

\section{Time evolution in spherical symmetry}\label{four}
In this section, we explain in detail how we perform fully non-linear evolutions of CNSs and summarize our numerical methods. We present results for the dynamics of CNS stars by analyzing their stability. We have considered screened  and descreened CNSs,  under  perturbations that trigger either oscillations or collapse to a BH.

\subsection{Evolution equations}
The fully non-linear evolution of CNS stars is followed in the Einstein frame, where the equations of motion for CNSs are given by the Einstein equations~\eqref{eq:einstein}; the conservation laws for the Einstein-frame baryon mass current [Eq.~\eqref{eq:consJ}] and stress-energy tensor [Eq.~\eqref{eq:consT}]; and the scalar field equation~\eqref{eq:kg}. We restrict our study to spherical symmetry and decompose the spacetime tensors into their space (radial) and time components.

We consider the following line element:
\begin{equation}\label{metric_ansatz}
ds^2 = -\alpha^2(t,r)dt^2+g_{rr}(t,r)dr^2+r^2g_{\theta\theta}(t,r)d\Omega^2~\,,
\end{equation}
where $\alpha(t,r)$ is the lapse function, $g_{rr}(t,r)$ and $g_{\theta\theta}(t,r)$ are positive metric functions, and
$d\Omega^2=d\theta^2+\sin^2\theta d\varphi^2$ is the solid angle element. These quantities are defined on each leaf $\Sigma_{t}$ of the spatial foliation, which has normal vector $n_{\mu}=(-\alpha,0)$ and extrinsic curvature $K_{ij} \equiv-\frac{1}{2}\mathcal{L}_{n}\gamma_{ij}$. Here, $\mathcal{L}_{n}$ is the Lie derivative along $n^{\mu}$ and $\gamma_{ij}$ is the metric induced on each leaf.

The Einstein equations \eqref{eq:einstein} are written as an evolution system by using the Z3 formulation in spherical symmetry~\cite{Alic:2007ev,Bona:2005pp}. We can express Eq.~\eqref{eq:einstein} as a first order  system by introducing first derivatives of the fields as independent variables, namely
\begin{eqnarray}
A_{r}=\frac{1}{\alpha}\partial_{r}\alpha \,,  \quad  {D_{rr}}^{r}=\frac{g^{rr}}{2}\partial_{r}g_{rr} \,, \quad  {D_{r\theta}}^{\theta}=\frac{g^{\theta\theta}}{2}\partial_{r}g_{\theta\theta} \,,\nonumber
\end{eqnarray}
and write the system of equations in the conservative form
\begin{equation}
\partial_{t}{\bf U} + \partial_{r} F( {\bf U} ) = \mathcal{S}( { \bf U}) \,,
\end{equation}
where ${\bf U}
=\{\alpha\,, g_{rr}\,, g_{\theta\theta}\,, {K_{r}}^{r}\,, {K_{\theta}}^{\theta}\,,A_{r}\,, {D_{rr}}^{r}\,, {D_{r\theta}}^{\theta}\,, Z_{r}\,\}$ is a vector containing the full set of evolution fields. In the Z3 formulation, the momentum constraint has been included in the evolution system by considering an additional vector $Z_{i}$ as an evolution field~\cite{Z42}. In fact, the $Z_{r}$ component is the time integral of the momentum constraint. In addition, $F( {\bf U} )$ is the radial flux and $\mathcal{S}( {\bf U})$ is a source term.
The evolution equations for the Z3 formulation can be found explicitly in  Ref.~\cite{ValdezAlvarado:2012xc}. A gauge condition for the lapse is required to close the system. We use the singularity-avoidance $1+\log$ slicing condition $\partial_{t}\alpha=-2\alpha\,\mathrm{tr}K,$ where $\mathrm{tr}K=K_{r}^{r}+2K^{\theta}\,_{\theta},$ see~\cite{BM}.

In addition, the equations of motion for the fluid [Eq.~\eqref{eq:consJ}-\eqref{eq:consT}] and for the scalar field [Eq.~\eqref{eq:kg}] are written in conservative form:
\begin{widetext}
\begin{align}
\partial_{t}(\zeta D) &= -\partial_{r}[\zeta D\alpha v^r] - \alpha\zeta D\left(\frac{2}{r} v^{r} -  \alpha_0(v^r\Phi+\sqrt{g^{rr}}\Pi)\right)~,\label{hydro1}\\
\partial_{t}(\zeta U)
&= -\partial_{r}[\zeta \alpha {S}^{r}] +
\alpha\zeta \left[ {S^{r}}_{r}{K^{r}}_{r}+2{S^{\theta}}_{\theta}{K^{\theta}}_{\theta}-S^{r}\left(A_{r} + \frac{2}{r}\right)- \alpha_0T\sqrt{g^{rr}}\Pi\frac{}{}\right]~,\label{hydro2}\\
\partial_{t}(\zeta S_r)
&=-\partial_{r}[\zeta \alpha {S^{r}}_{r}] + \alpha\zeta  \left[{{S}^{r}}_{r}\left({D_{rr}}^{r}-\frac{2}{r}\right)+2{S^{\theta}}_{\theta}\left({D_{r\theta}}^{\theta}+\frac{1}{r}\right)-U A_{r}+  \alpha_0T\Phi\frac{}{}\right]\label{hydro3}\,,\\
\partial_{t}\phi &= \frac{\alpha}{\sqrt{g_{rr}}}\,\Pi~,\\
\partial_{t}\Phi &= \partial_{r}\left[\frac{\alpha}{\sqrt{g_{rr}}}\,\Pi\right]~,\\
\partial_{t}\Pi &= \partial_{r}\left[\frac{\alpha}{\sqrt{g_{rr}}}\Phi\right] + \frac{\alpha}{\sqrt{g_{rr}}}\left( 2\left({D_{r\theta}}^{\theta}+\frac{1}{r}\right)\Phi + 2\sqrt{g_{rr}}{K_{\theta}}^{\theta}\Pi -g_{rr}\frac{dV_{\text{eff}}}{d\phi}\right)\,,
\end{align}
\end{widetext}
where $\zeta  =\sqrt{g_{rr}}g_{\theta\theta}$ and
\begin{eqnarray}
\Phi&=&\partial_{r}\phi \,, \qquad\Pi=\frac{\sqrt{g_{rr}}}{\alpha}\partial_{t}\phi \,.
\end{eqnarray}
Note that Eqs.~\eqref{hydro1}-\eqref{hydro3} are given in terms of the  \textit{conserved} quantities $\{D,U,S_{r}\}$, which are defined in terms of the physical (or \textit{primitive}) variables, i.e.:  fluid pressure $p$, rest-mass density $\rho$, specific internal energy $\xi$~\footnote{Note that this quantity must not be confused with the (total) energy density, $\epsilon$. The connection between the too is given by the relation $\xi=\epsilon/\rho-1$, from which it becomes clear that $\xi$ is adimensional.}, radial velocity of the fluid $v^{r}$, and the enthalpy of the fluid, $h \equiv \rho (1 + \xi) + p.$ The conserved quantities are explicitly defined as follows:
\begin{eqnarray}
D &=& \rho W~,~~U =  h W^2 - p ~,~~ {S}_r =  h W^2 v_r~,\label{contoprim}\\
{S_{r}}^{r} &=& hW^{2}v_{r}v^{r} + p~,~~{S_{\theta}}^{\theta} =  p~,~~ T=-h+4p~,
\end{eqnarray}
with  $W^2= 1/(1-v_{r}v^{r})$ the  Lorentz factor, and  ${S_{r}}^{r}$ and ${S_{\theta}}^{\theta}$ the spatial projections of the stress energy tensor of the fluid
in the Einstein frame. Finally, to recover the physical fields  $\{\rho,\xi,p,v^{r}\}$ during the evolution, the algebraic relation~\eqref{contoprim} has to be inverted, which involves solving a nonlinear equation at each time-step. During this process, we employ an ideal-gas EoS $P= (\Gamma -1)\rho\xi$ (see Appendix B in Ref.~\cite{suspalen}), with the appropriate $\Gamma$ depending on the CNS simulation, as explained in Sec.~\ref{three}.

\subsection{Implementation}

The one-dimensional (1D) numerical code used in this work is an extension of the one presented in Ref.~\cite{suspalen} for fully non-linear simulations of fermion-boson stars, and used in Refs.~\cite{Alic:2007ev,Bernal:2009zy,Raposo:2018rjn,terHaar:2020xxb} to study the dynamics of BHs, boson stars, anisotropic stars and NSs with kinetic screening mechanism.
As initial data, we use the static CNS solutions discussed in Sec.~\ref{three}, transformed
from the areal coordinates of Eq.~\eqref{eq:ansatz} to maximal isotropic coordinates, in which the line element is given by
\begin{eqnarray}
ds^{2} = -\alpha^{2}(r) dt^{2} + \psi^{4}(r)(dr^{2} + r^{2}d\Omega^{2})~,
\label{metric_conformal}
\end{eqnarray}
being $\psi$  the conformal factor.

We have used a high-resolution shock-capturing finite-difference (HRSC) scheme, described in Ref.~\cite{Alic:2007ev}, to discretize the spacetime, the scalar field and the fluid matter fields. In particular, this method can be viewed as a fourth-order finite difference scheme plus third-order adaptive dissipation. The dissipation coefficient is given by the maximum propagation speed at each grid point. The method of lines is used to perform the time evolution through a third-order accurate strong stability preserving Runge-Kutta integration scheme, with a Courant factor of $\Delta t/\Delta r = 0.25$ (in code units, $G=c=M_{\odot}=1$), so that the Courant-Friedrichs-Levy condition imposed by the principal part of the system of equations is satisfied. Most of the simulations presented in this work have been performed with spatial resolutions of $\Delta r =\{ 0.005, 0.0025, 0.00125\}~GM_{\odot}$, in a domain with outer boundary located between $r= 500~GM_{\odot}$ and $r= 1000~GM_{\odot}$. We have verified convergence of results with increasing resolution as well as their robustness against changes in the position of the outer boundary. We use maximally dissipative boundary conditions for the spacetime variables, and outgoing boundary conditions for the scalar field and for the fluid matter fields.

\subsection{Screened and descreened CNSs}
To test the stability of CNSs, we have first evolved the initial data described in Sec.~\ref{three},
subjected only to the small perturbations given by truncation errors. In addition, we have tested the migration of CNSs from the unstable to the stable branch of solutions. In the subsections below we report and discuss examples of such tests. Finally, we discuss the results from simulations of gravitational collapse to a BH. All results shown in this section have been produced for the parameter choice $(\Lambda,\teinf)=(175$ GeV, $6.5\times 10^{10}$ g/cm$^3$).

\subsubsection{Stability}

\begin{figure}[ht]
\hspace{-1cm}
\includegraphics[width=0.5\textwidth]{ 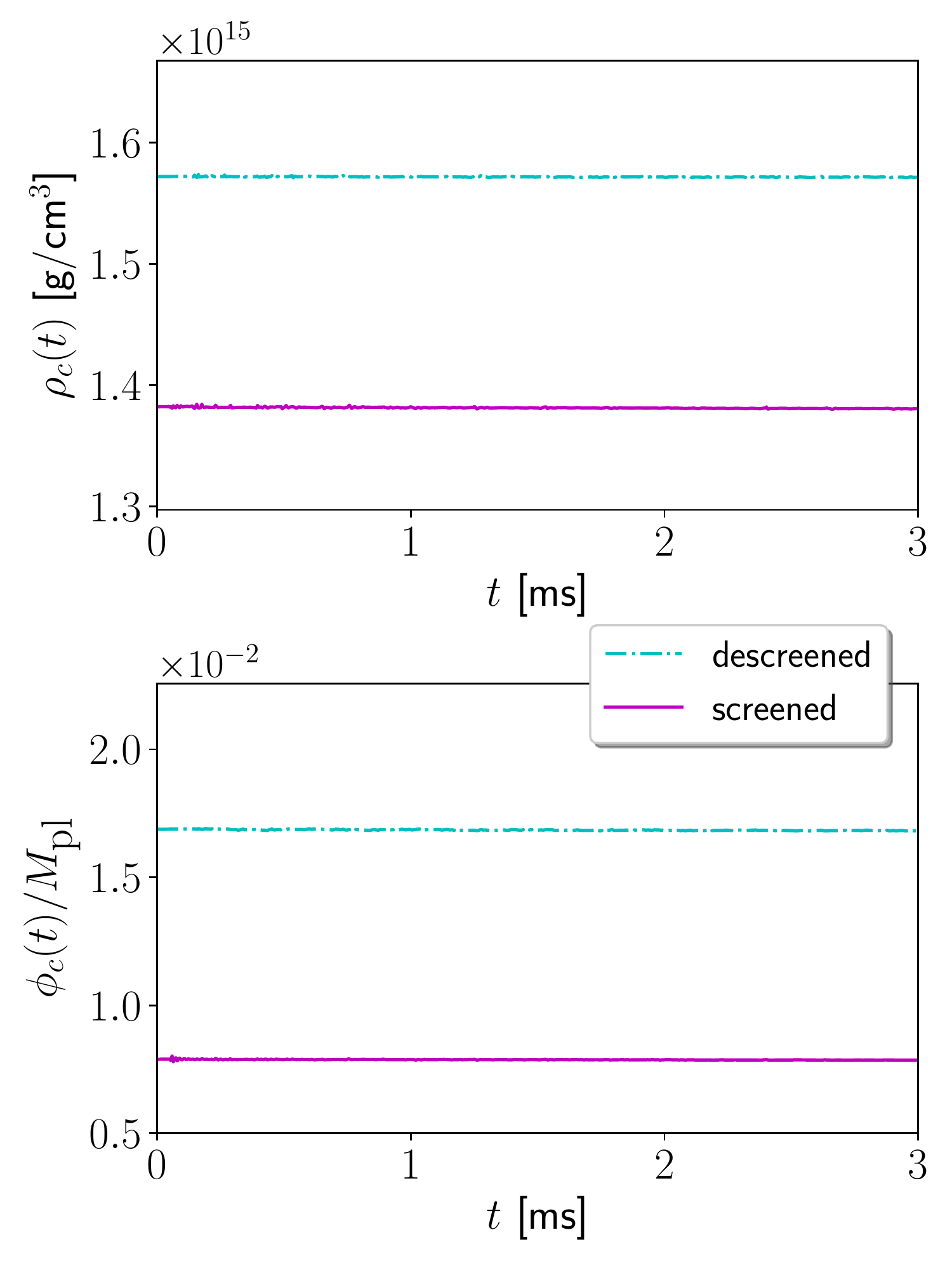}
\caption{{\em Stability of CNS.} Top panel: $\rho_c(t)\equiv\tilde{\rho}(t,r=0)$ (Jordan frame) vs time. Bottom panel: $\phi_c(t)\equiv\phi(t,r=0)$ vs time. The (magenta) solid line corresponds to a screened CNS with initial central density $\rho_c\simeq1.38\times10^{15}$ g/cm$^{3}$. The (cyan) dash-dotted line corresponds to a descreened CNS with initial central density $\rho_c\simeq1.57\times10^{15}$ g/cm$^{3}$. The matter and chameleon field configurations are stable against small perturbations given by truncation errors.}
\label{fig:stab}
\end{figure}

In Fig.~\ref{fig:stab} we show the time evolution of the central density (upper panel) and central scalar field  (bottom panel) for two CNSs, one with complete screening (solid magenta line) and one with partial descreening in the core (dash-dotted cyan line). The first star has a lighter gravitational mass $M=1.72 M_{\odot}$ and initial (Jordan-frame) central density $\rho_c\simeq1.38\times10^{15}$ g/cm$^{3}$. The descreened star is heavier, with a mass of $M=1.84 M_{\odot}$ and initial central density $\rho_c\simeq1.57\times10^{15}$ g/cm$^{3}$.
The simulations were conducted on a grid that extends up to $r=1000~GM_{\odot}$ with a spacing as fine as $\Delta r=0.0025~GM_{\odot}$ for the screened star. For simulations of the descreened star, however, we have doubled the number of points of our spatial grid, which correspond to  $\Delta r = 0.00125~GM_{\odot}$.
We have observed that simulations of descreened stars are more challenging, as higher resolutions are typically needed to keep the numerical dissipation under control during the evolution. We interpret this technical issue as stemming again from the separation between stellar and cosmological scales.

Both stars were evolved in time with no other perturbation but the one introduced by truncation errors: their stability is manifest in Fig.~\ref{fig:stab}, which shows that the central density and central scalar field remain constant over time.

\subsubsection{Migration}
\begin{figure}[h!t]
\centering
\hspace{-1cm}
\includegraphics[width=0.53\textwidth]{ 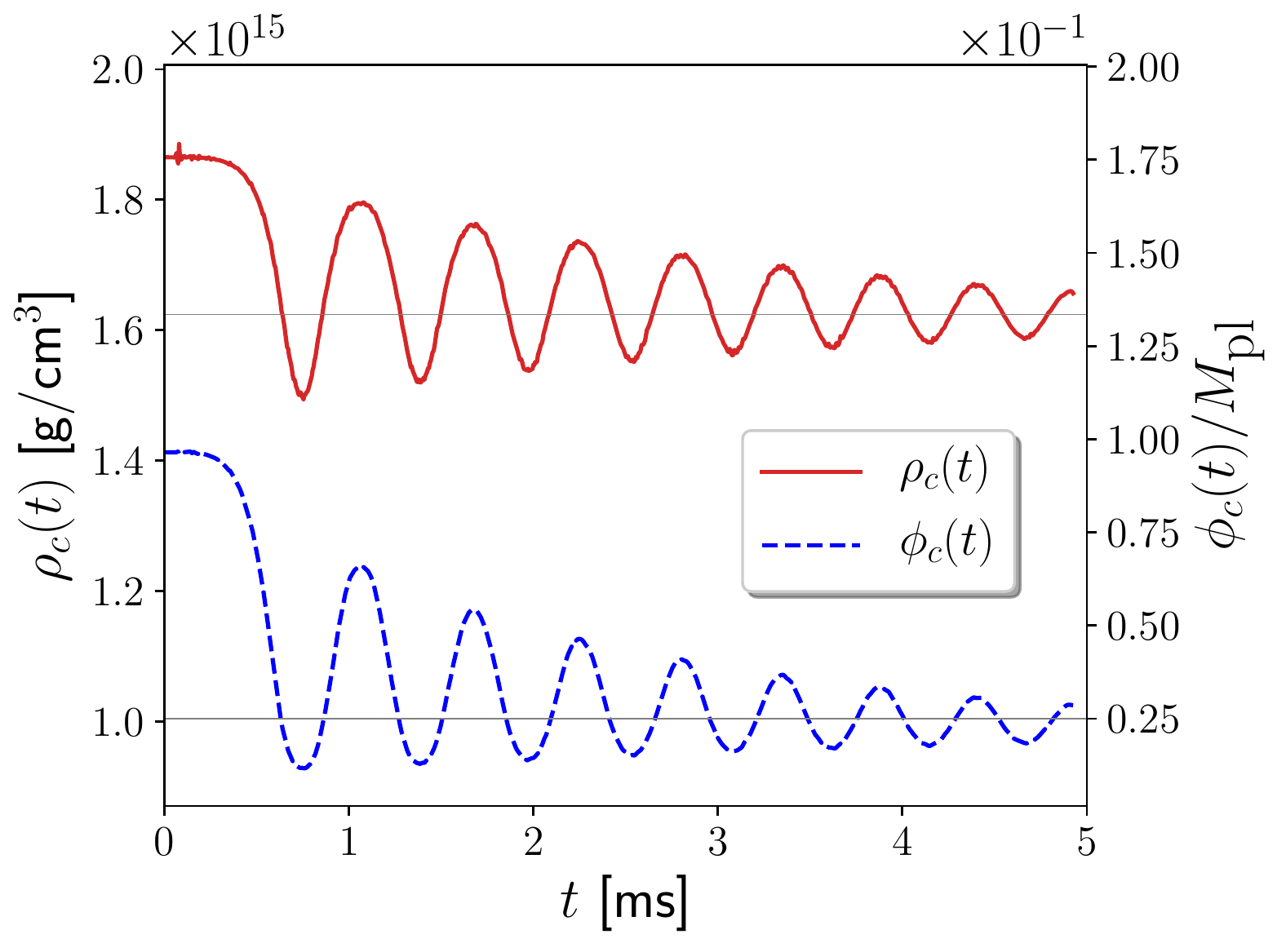}
\caption{ {\em{Migration test}}. The time evolution  of  the (Jordan-frame) central density $\rho_c(t)\equiv\tilde{\rho}(t,r=0)$ (red solid line)  and  central scalar field $\phi_c(t)\equiv\phi(t,r=0)$ (blue dashed line), for an unstable and partially descreened CNS with baryon mass $M_{bar}\simeq 1.88M_{\odot}$ and $\rho_c\simeq1.87\times10^{15}$ g/cm$^{3}$. The star is expanding in volume and relaxes through large dampened oscillations to a stable descreened CNS with the same mass, but lower central density $\rho_c\simeq1.66\times10^{15}$ g/cm$^{3}$.}
\label{fig:migration}
\end{figure}

The migration test is a standard diagnostics tool utilized in GR to characterize the (in)stability of NS solutions (e.g., see~\cite{Font2000,Radice_2010,PhysRevD.84.044012}): depending on the initial perturbation~\cite{Font:2001ew}, solutions that lie on the unstable branch (i.e. to the left of the maximum mass configuration in mass-radius plots such as Fig.~\ref{fig:MassRadius}) can either collapse to a BH or undergo a series of wide oscillations and migrate towards a solution on the stable branch (with approximately the same baryon mass). In our simulations, migration of highly compact and unstable CNSs is induced via small perturbations given by the truncation error. An example of migration is given in Fig.~\ref{fig:migration}, where a star with initial central density $\rho_c\simeq1.87\times10^{15}$ g/cm$^{3}$ can be seen undergoing large dampened oscillations, which eventually relax it to a stable descreened star with approximately the same baryon mass $M_{bar}= 1.88M_{\odot}$ (modulo a small loss due to numerical dissipation) and central density $\rho_c\simeq1.66\times10^{15}$ g/cm$^{3}$.
\subsubsection{Spherical collapse}
We have conducted simulations of spherical collapse to BHs, which are another standard benchmark for numerical relativity simulations of NSs. The collapse has been induced by an initial pressure gradient up to ten percent. We illustrate the results of this test by discussing the case of a collapsing descreened CNS with
gravitational  mass $M=1.89M_{\odot}$ and initial central density $\rho_c=1.70\times10^{15}$ g/cm$^{3}$. In  Fig.~\ref{fig:rho,alpha,phi},  we show the time evolution of the density, chameleon field and lapse at the center of the collapsing star. As matter collapses to the center of the star, the density and pressure in the core grow, pushing the chameleon field down its effective potential (i.e. to higher values).
This is counterintuitive, as the minimum of the effective potential \eqref{eq:effpot}
moves to smaller $\phi$ when the density increases, as long as the star remains non-relativistic. However,
this behavior breaks down when the configuration becomes relativistic, as a result of the change of sign of $T_m$.
Indeed, for $T_m>0$ the effective potential has no minimum, and the scalar field rolls down to larger and larger values.

The lapse decreases to zero and, as a consequence of the $1+\log$ slicing coordinate choice that we employ, the time evolution of matter in the collapsing core is effectively frozen. In Fig.~\ref{fig:collapse}, we show time snapshots of the radial profile of the lapse and chameleon scalar field. Inside the star, as the lapse goes to zero an apparent horizon (black dots) forms and slowly expands, until it eventually engulfs the whole matter content. While the chameleon field inside  the apparent horizon grows as a result of the (runaway) effective potential, outside the horizon it slowly relaxes to the exterior configuration minimizing the effective potential in the presence of an atmosphere. No instabilities develop during collapse
outside the apparent horizon, and the end state  is therefore a BH with a trivial scalar field solution.
\begin{figure}[ht]
\centering
\includegraphics[width=0.45\textwidth]{ 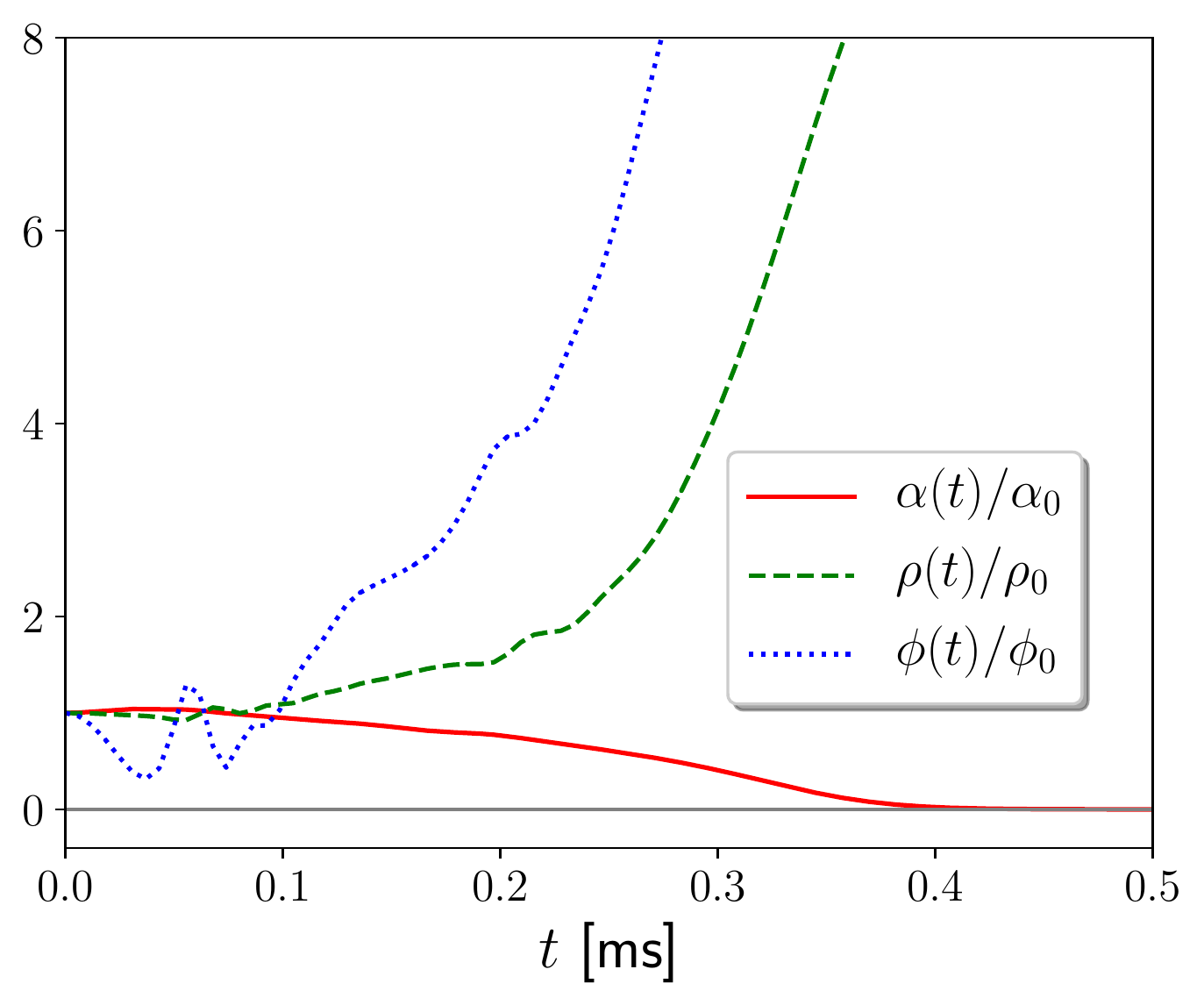}
\caption{ {\em{Collapse of a descreened CNS}}. Time evolution of the lapse (solid red line), density (dashed green line) and chameleon field (dotted blue line) at the center of a collapsing descreened star with gravitational mass $M=1.88M_{\odot}$, normalized by their initial value. As matter collapses towards the center, the  density  chameleon field increase. The lapse function decreases to small values close to zero.}
\label{fig:rho,alpha,phi}
\end{figure}
\begin{figure}[ht]
\centering
\includegraphics[width=0.5\textwidth]{ 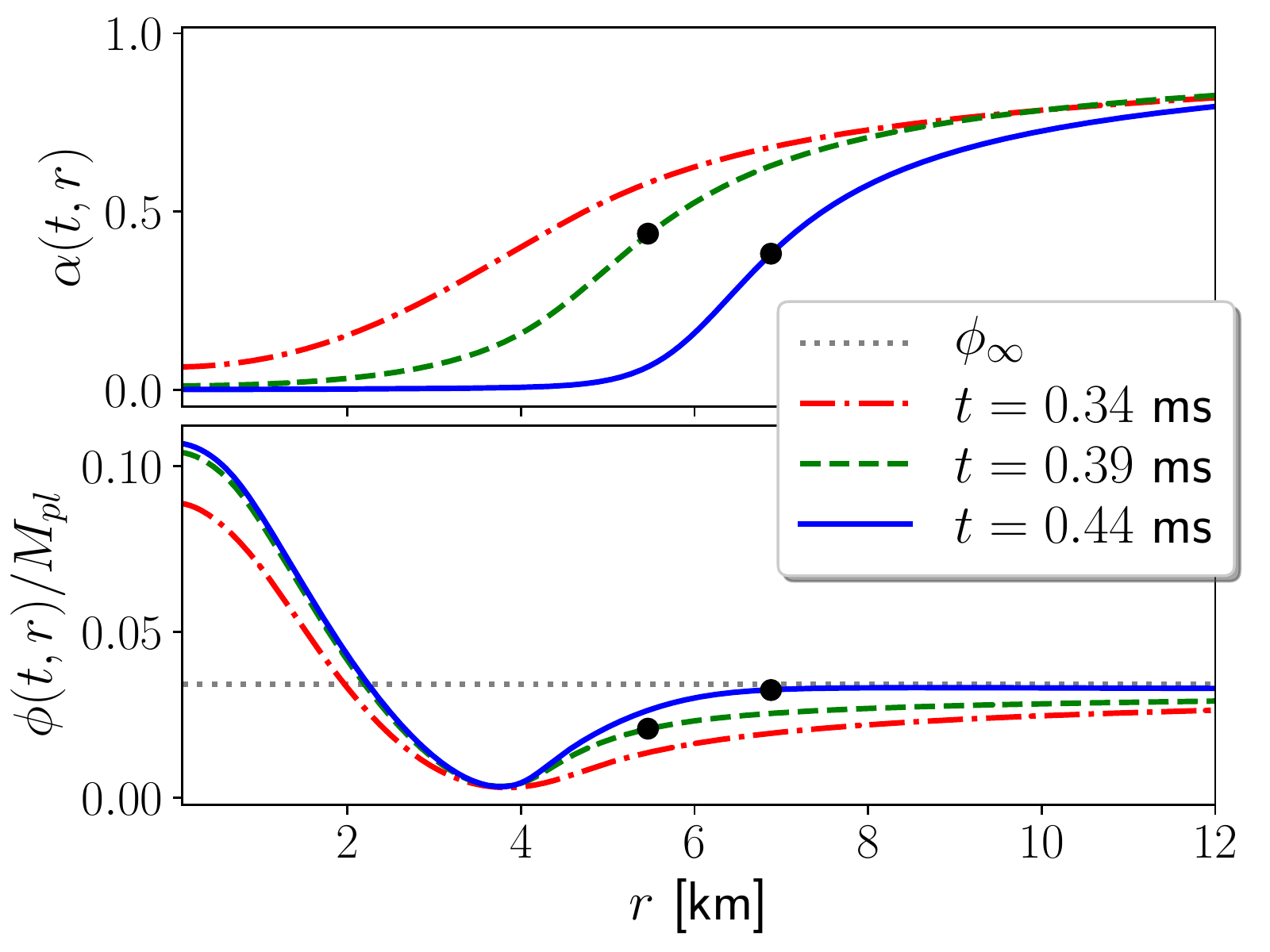}
\caption{{\em{Collapse of a descreened CNS.}}
Snapshots of the radial profile of the lapse function (top panel) and scalar field (bottom panel) for the same collapsing star as in Fig.~\ref{fig:rho,alpha,phi}.
Black dots indicate the position of the apparent horizon.}
\label{fig:collapse}
\end{figure}
\section{Radial oscillations}\label{five}
In this section, we analyze the spectrum of the radial oscillations of spherically symmetric CNSs, and compare to the oscillation spectrum of NSs with similar gravitational masses in GR. The CNSs have been produced with the parameter choice $(\Lambda,\teinf)=(175$ GeV, $6.5\times 10^{10}$ g/cm$^3$). As a first step, we test the accuracy of our code by producing a NS in GR, with gravitational mass $M=1.4 M_{\odot}$ and EoS defined by $\Gamma=2$ and $K=100~G^3M_{\odot}^2c^{-4}$. From sufficiently long simulations, the frequencies of the characteristic radial oscillations (induced by truncation errors) have been extracted and compared with the ones estimated in~\cite{Font:2001ew} from an independent three-dimensional (3D) code. The results, summarized in Table~\ref{tab:GR2K100}, are an indicator of the accuracy of our frequency estimates.

\begin{center}
\begin{table}[ht]
\begin{tabular}{|c|c|c|c|}
\hline
mode & 1D code (kHz) & 3D code (kHz) & Rel. diff. ($\%$) \\
\hline\hline
F  & 1.443 &  1.450 & 0.6 \\
H1 & 3.952 &  3.958 & 0.2 \\
H2 & 5.902 &  5.935 & 0.6 \\
H3 & 7.763 &  7.812 & 0.6 \\
\hline
\end{tabular}
\caption{{\em{Radial oscillations frequencies of a NS in GR.}} Comparison between estimates with our 1D code vs an independent 3D code~\cite{Font:2001ew}.}
\label{tab:GR2K100}
\end{table}
\end{center}

\begin{figure}[ht]
\centering
\includegraphics[width=0.5\textwidth]{ 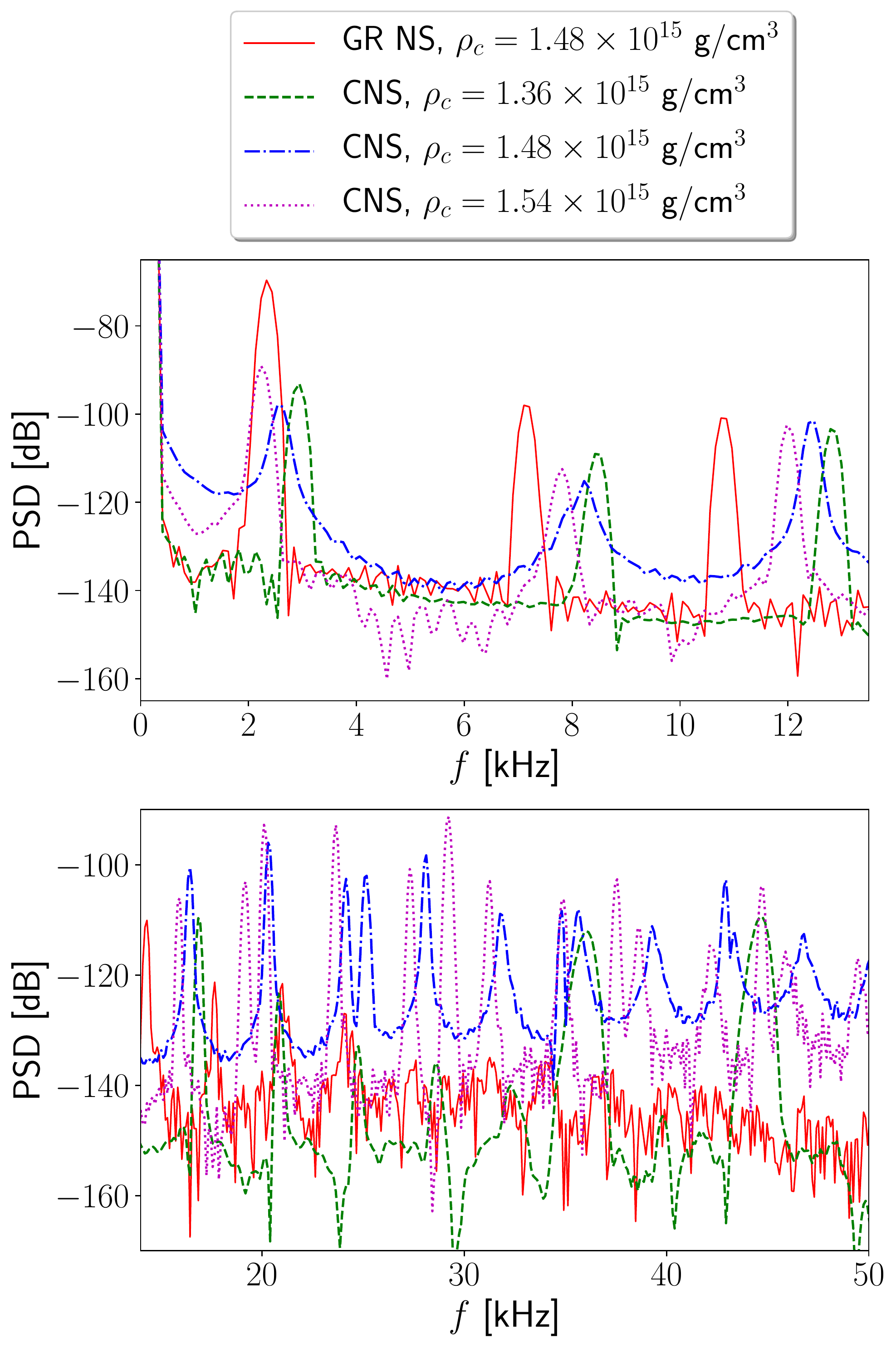}
\caption{{\em{Spectra of radial oscillations}}. The two plots show the PSDs of the radial modes extracted from the time evolution of the central rest-mass density for three CNSs and one GR NS. Top panel: F, H1 and H2 modes. Bottom panel: higher overtones, $H_N$ with $N>2$, and the new family of scalar modes ($F_s$ and higher overtones). The results shown in this plot are valid for the parameter choice $(\Lambda,\teinf)=(175$ GeV, $6.5\times 10^{10}$ g/cm$^3$).}
\label{fig:qnms}
\end{figure}

From long-term simulations of several CNSs with central densities in the range $\rho_{c}=(0.96$ -- $1.67)\times10^{15}$ g/cm$^{3}$, we have then computed the power spectral density (PSD) of the density perturbations and extracted the peak frequency of the fundamental radial mode ($F$) and its higher overtones ($H_N$, with $N=1,2,...$). As a reference, we have also evolved spherical NSs produced in GR with comparable gravitational masses, using the same EoS ($\Gamma=3$).
The presence of the chameleon field coupled to matter inside the star has multiple effects on the spectrum. The first, which can be observed in Fig.~\ref{fig:qnms}, consists in a modification of the relation between the peak frequencies and the properties of the stars.

\begin{figure}[ht]
\centering
\includegraphics[width=0.5\textwidth]{ 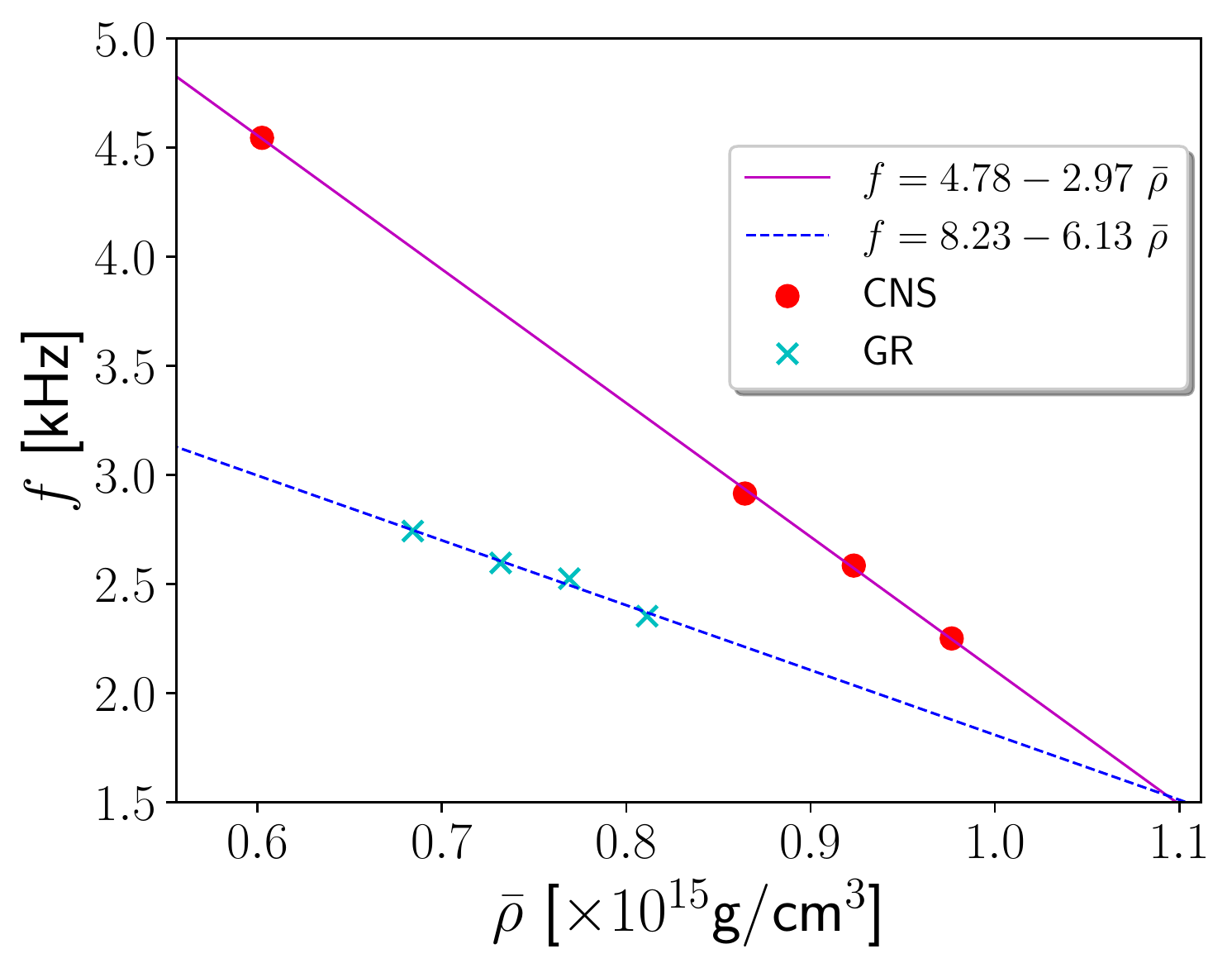}
\caption{{\em{$F$-mode frequency vs average density}}. Linear fits of the fundamental mode ($F$) frequency as a function of the average density ($\bar{\rho}$) for NSs respectively in GR (blue dotted line, cyan cross tokens) and in chameleon gravity (magenta dash-dotted line, red round tokens). The results shown in this plot are valid for the parameter choice $(\Lambda,\teinf)=(175$ GeV, $6.5\times 10^{10}$ g/cm$^3$).}
\label{fig:rhoVSfreq}
\end{figure}

Radial oscillations of NSs in GR have been studied extensively in the past. For instance, it is known that non-relativistic homogeneous stars feature a fundamental mode frequency, $F$, that is proportional to the (constant) rest-mass density~\cite{1983bhwd.book.....S}. This relation
 is more complicated in the relativistic regime, and the result for non-relativistic homogeneous stars only holds  approximately at low densities~\cite{1992A&A...260..250V,Kokkotas:RadOsc}. In order to quantify the difference between spectra in GR and chameleon gravity, we have fitted the relation between the $F$-mode frequency and the average density, $\bar{\rho}\equiv (4\pi/3)^{-1} M/R_{star}^3$, in either theory. We present the result of the comparison in Fig.~\ref{fig:rhoVSfreq}.

The additional scalar degree of freedom of ST theories can also produce a new family of characteristic oscillations inside NSs. These scalar radial modes
correspond to monopole GW emission. Indeed, in the spectra of CNSs, we observe several high-frequency peaks that do not have any correspondence in the GR power spectra (see Fig.~\ref{fig:qnms}, bottom panel). We interpret these peaks as due to the chameleon field oscillations.
The fundamental (massive) scalar mode of oscillation has a frequency, $F_s\equiv m_{\textrm{eff}}/2\pi$, that is of order of the inverse of the Compton wavelength: the larger the mass, the larger the corresponding frequency (see e.g., Fig. 2 in ~\cite{Blazquez-Salcedo:2020ibb}). For $\Lambda=175$ GeV, the chameleon field inside objects as dense as NSs acquires a very large mass~\eqref{eq:mass}, which yields frequencies $F_s\sim O(10)$ kHz.
This is indeed the correct order of magnitude for the frequencies of the new family of modes that we observe.
For $\Lambda\simeq2.4$ meV, one can check that $F_s\gg$ kHz, because the chameleon  acquires even larger masses inside relativistic stars. These modes are hardly  excited and are unobservable with GW detectors.

Regarding the shift in the peak frequencies, note that such effect is present even in CNSs with a screened interior (see e.g. the $\rho_c\simeq1.38\times10^{15}$ g/cm$^{3}$ configuration in Fig.~\ref{fig:qnms}). Like in the case of the mass-radius relation (c.f. Fig.~\ref{fig:MassRadius} and related discussion), we expect deviations from GR in the spectrum of oscillations to disappear
in the limit $\Lambda\rightarrow 2.4$ meV for screened stars, while they could survive for descreened CNSs. As we will see in the following, however,
these effects are likely outside the reach of ground-based gravitational interferometers.

\section{Scalar radiation}\label{six}
In this section, we investigate the characteristic GW output of CNSs,
focusing on detectability with current and future detectors. To this end, for each signal produced with our simulations
we estimate the signal-to-noise ratio (SNR) as~\cite{Maggiore2007}
\begin{equation}\label{eq:snr}
\textrm{SNR}^2\equiv\int_0^{\infty}\frac{4|\tilde{h}(f)|^2}{S_n(f)}df\,,
\end{equation}
where $\tilde{h}(f)$ is the strain signal  in the frequency domain and $S_n(f)$ is the one-sided noise power spectral density of the detector. As a reference, we compare the simulated signals with the design sensitivity curves of the Advanced Laser Interferometer Gravitational-Wave Observatory (Advanced LIGO)~\cite{TheLIGOScientific:2014jea,aligo}\footnote{For the sensitivity we refer to the zero detuning, high power configuration.}, Einstein Telescope (ET)~\cite{Hild:2010id} and Laser Interferometer Space Antenna (LISA)~\cite{2017arXiv170200786A,lisa}.
The geometry of the detector is encoded in the pattern functions, $F_+, F_{\times}, F_0$, which
are different in the case of a tensor wave ($\tilde{h}(f)=F_+\tilde{h}_+(f) + F_{\times} \tilde{h}_{\times}(f)$)
and for a scalar wave (breathing mode, $\tilde{h}(f)=F_0\tilde{h}_0(f)$). For simplicity,
we will assume optimal detector orientation~\cite{Nishizawa:2009bf,Yunes:2013dva,Gerosa:2016fri} in our calculations, i.e. $F_0=1/2$.

The effect of GWs on the detector  is encoded in the Newman-Penrose curvature scalars~\cite{Newman:1961qr}. The latter can be obtained by projecting the
Riemann tensor onto a null tetrad basis $\left(k,l,m,\bar{m}\right)$ adapted to the wavefronts.
In particular, the scalar mode is encoded in
$\Phi_{22}=-R_{lml\bar{m}}$ (evaluated in the Jordan frame)~\cite{PhysRevD.8.3308}.
This quantity can be computed from our simulations
(which are performed in the Einstein frame)
via
\begin{eqnarray}
\Phi_{22} = A(\phi)^{-2}& &\left(\Phi_{22}^E + l^al^b\nabla_a\nabla_b\log A(\phi)\right.\nonumber\\
&& \left. - (l^a\nabla_a\log A(\phi))^2\right)~\,,
\end{eqnarray}
where $\Phi_{22}^E$ is the same Newman-Penrose scalar in the Einstein frame.
Since in that frame the ST theories that we consider simply reduce (in vacuum) to
GR with a minimally coupled scalar field, we can conclude that $\Phi_{22}^E\simeq0$,
and the only significant contribution comes from the oscillating chameleon field,
i.e.~\cite{Barausse:2012da}
\begin{equation}\label{eq:phi22}
\Phi_{22}\simeq 2\alpha_0 \partial_t^2\phi+O\left(\frac{1}{r^2}\right)\,,
\end{equation}
in deriving which we have used $\partial_t^2\varphi\sim O(1/r)$
and neglected terms decaying as $1/r^2$ or faster.
In practice, $\Phi_{22}$ is computed from our simulations by evaluating~\eqref{eq:phi22} at an extraction radius placed sufficiently far away from the star, $r_{ext}\gg R_{star}$. At the same time, the extraction radius must be far from the cosmological horizon, $r_{ext}\ll r_{cosmo}$, in an intermediate region where geometric effects from the de Sitter asymptotics are negligible and the spacetime is approximately flat. In addition to the spacetime flatness requirement, the extraction radius must also be chosen to satisfy $r_{ext}\gg \lambda_c=1/m_{\infty}$.
By combining all the requirements listed above one obtains the radiation zone condition~\cite{Sperhake:2017itk}, $\lambda_c\ll r_{ext}\ll r_{cosmo}$. (Note that one typically has $\lambda_c\gg R_{star}$.)
Because of the rather large effective cosmological constant, in our simulations the wave zone requirements are met only in a rather tight range of the isotropic radius coordinate (e.g. $r_{ext}=50$--$100~GM_{\odot}$ for $\Lambda=175$ GeV). We have checked that our results are robust with respect to variations of the extraction radius in this range and to the position of the outer boundary of our simulations, which we place sufficiently far from the extraction point, at distances typically larger than $500 ~GM_{\odot}$.

The signal is produced as a function of the retarded time, $t_{ret}\approx t - r_*$, defined in terms of the Schwarzschild-de Sitter tortoise coordinate, $r_*\equiv\int dr/f(r)$. This approximate prescription works well for our purposes, even though more involved expressions
can be employed~\cite{Boyle:2009vi,Bishop:2016lgv}.
We finally reconstruct the scalar strain in two independent ways. In the first method, with a Fast Fourier Transform algorithm we compute the frequency-domain Newmann-Penrose scalar $\tilde{\Phi}_{22}(f)$, from which we reconstruct the scalar strain $h_s$ (with $\Phi_{22}\equiv \partial_t^2 h_s$) with the following filter in the frequency domain:
\begin{align}
\tilde{h}_s(f)=
\left\lbrace
\begin{array}{lc}
-\frac{1}{(2\pi f)^2}\tilde{\Phi}_{22}(f) & f> f_0 \\
-\frac{(2\pi f)^2}{(2\pi f_0)^4}\tilde{\Phi}_{22}(f) & f \leq f_0
\end{array}
\right.\,,
\end{align}
inspired by~\cite{2011CQGra..28s5015R,Bishop:2016lgv} with the addition of a factor $\sim(f/f_0)^2$ suppressing unphysical low-frequency noise.
The frequency cutoff, $f_0$, is chosen according to the lowest physical frequency of the system. In practice, for simulations of oscillating stars we fix this to be of the order of the fundamental radial mode, $F$, since under
this threshold there is no stellar  mode that can source the scalar radiation. Instead, the gravitational collapse
produces what is sometimes referred to as an ``inverse chirp''~\cite{Gerosa:2016fri,Sperhake:2017itk,Rosca-Mead:2020ehn}: the GW burst excites lower and lower frequencies as the matter collapses. In this case, the mass of the chameleon field in the exterior introduces a natural cutoff frequency, $f_{\infty}\equiv m_{\infty}/2\pi$, as the propagation of modes with lower frequencies, $f\lesssim f_{\infty}$, is exponentially suppressed.
As a test, we checked the robustness of our results by varying the cutoff frequency down to the
lowest resolvable frequency in our simulations, $f_0\simeq 1/T$, where $T$ is the total simulation time.
The second method consists in computing the strain of the scalar monopole radiation directly from the formula~\cite{Gerosa:2016fri,Sperhake:2017itk,Rosca-Mead:2020ehn}
\begin{equation}
h_s=2\alpha_0(\phi-\phi_{\infty})\,,
\end{equation}
which can be derived by combining $\Phi_{22}=\partial_t^2h_s$ and Eq.~\eqref{eq:phi22}, which is approximately valid in the ``wave zone'' defined earlier. The agreement of the results obtained with the two methods confirms the robustness of our conclusions.

\subsection{Oscillating CNSs}

Oscillations in the CNSs were induced by an initial perturbation in the specific internal energy~(see sec.~\ref{four}.A), $\delta\xi(r)=\delta\xi_0\cos(\sigma r)\exp(-r^2/\sigma^2)$, with $\sigma=5~GM_{\odot}$ and $\delta\xi_0=\{10^{-6},10^{-5},10^{-4},10^{-3},4\times 10^{-3}\}$.
We have compared CNSs with different masses, the lighter one having $M=1.02 M_{\odot}$ and belonging to the screened branch of solutions, while the heavier, $M=1.84 M_{\odot}$, belongs to the branch with partial descreening. Here we take  $(\Lambda,\teinf)=(175$ GeV, $6.5\times 10^{10}$ g/cm$^3$).

\begin{figure}[th!]
\centering
\includegraphics[width=0.5\textwidth]{ 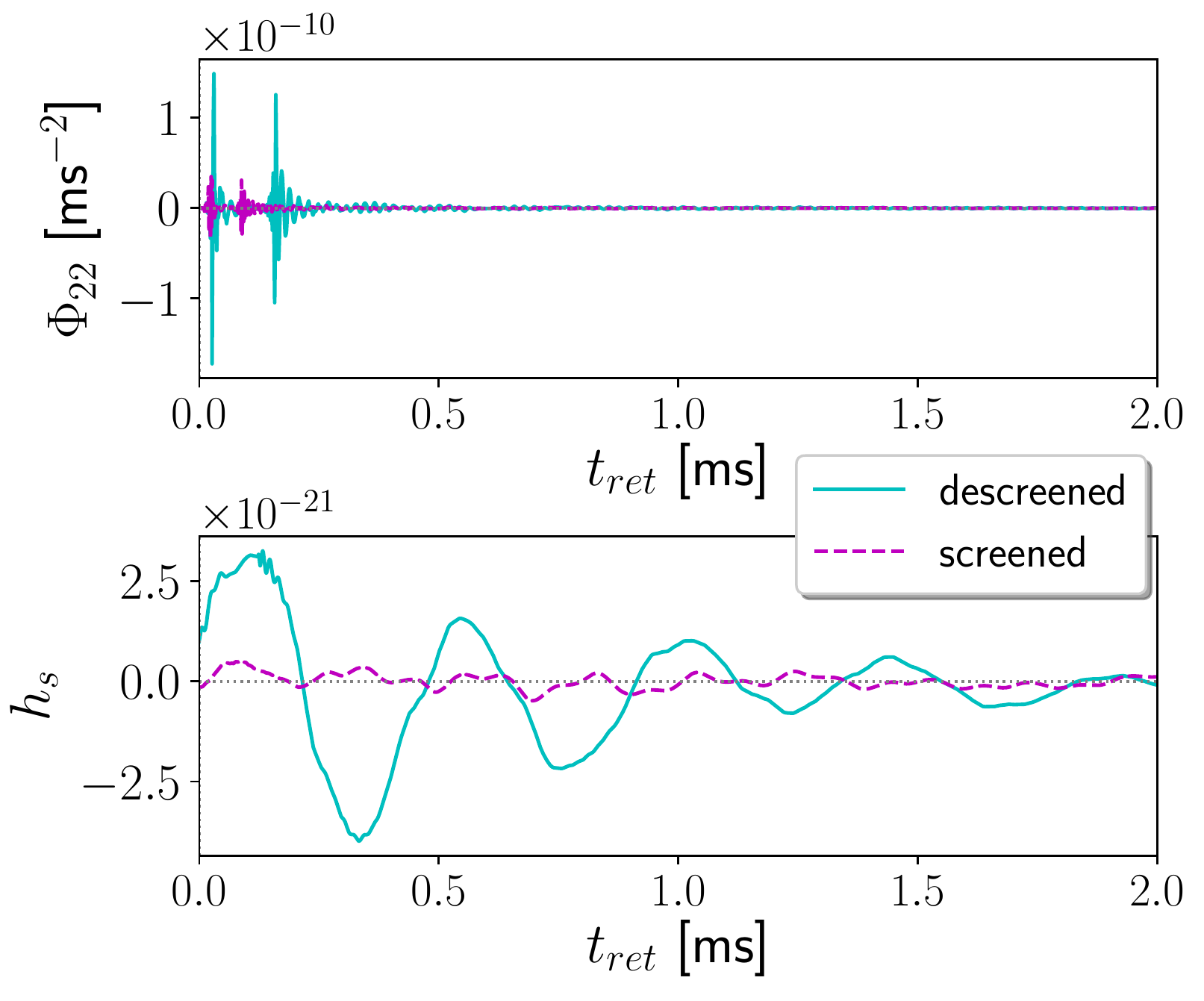}
\caption{ {\em{Scalar radiation from oscillating stars.}} Top panel: $\Phi_{22}$ vs retarded time. Bottom panel: strain amplitude vs retarded time. The results shown in these plots have been obtained for  $(\Lambda,\teinf)=(175$ GeV, $6.5\times 10^{10}$ g/cm$^3$). The scalar radiation is extracted from simulations of oscillating CNSs respectively with (dashed magenta lines) and without (continuous cyan lines) screening in the interior. The gravitational masses of the stars are, respectively, $M=1.02 M_{\odot}$ (screened CNS) and $M=1.84 M_{\odot}$ (descreened CNS). The distance of the detector from the source is set to $D_L=10$ kpc. To trigger the oscillations, an initial perturbation with amplitude $\delta\xi_0=10^{-6}$ is employed. }
\label{fig:gwosc}
\end{figure}

Let us first assess the effectiveness of the screening mechanism at suppressing the scalar radiation emitted by CNSs. In Fig.~\ref{fig:gwosc} we plot the
monopole GW signal sourced by an oscillating star at a luminosity distance of $D_L=10$ kpc. One can  observe that both the $\Phi_{22}$ curvature scalar and the strain amplitude $h_s$, respectively in the top and bottom panel, are suppressed (by a factor $\sim O(10)$) when the screening mechanism is active inside the star.
\begin{figure}[th!]
\centering
\includegraphics[width=0.5\textwidth]{ 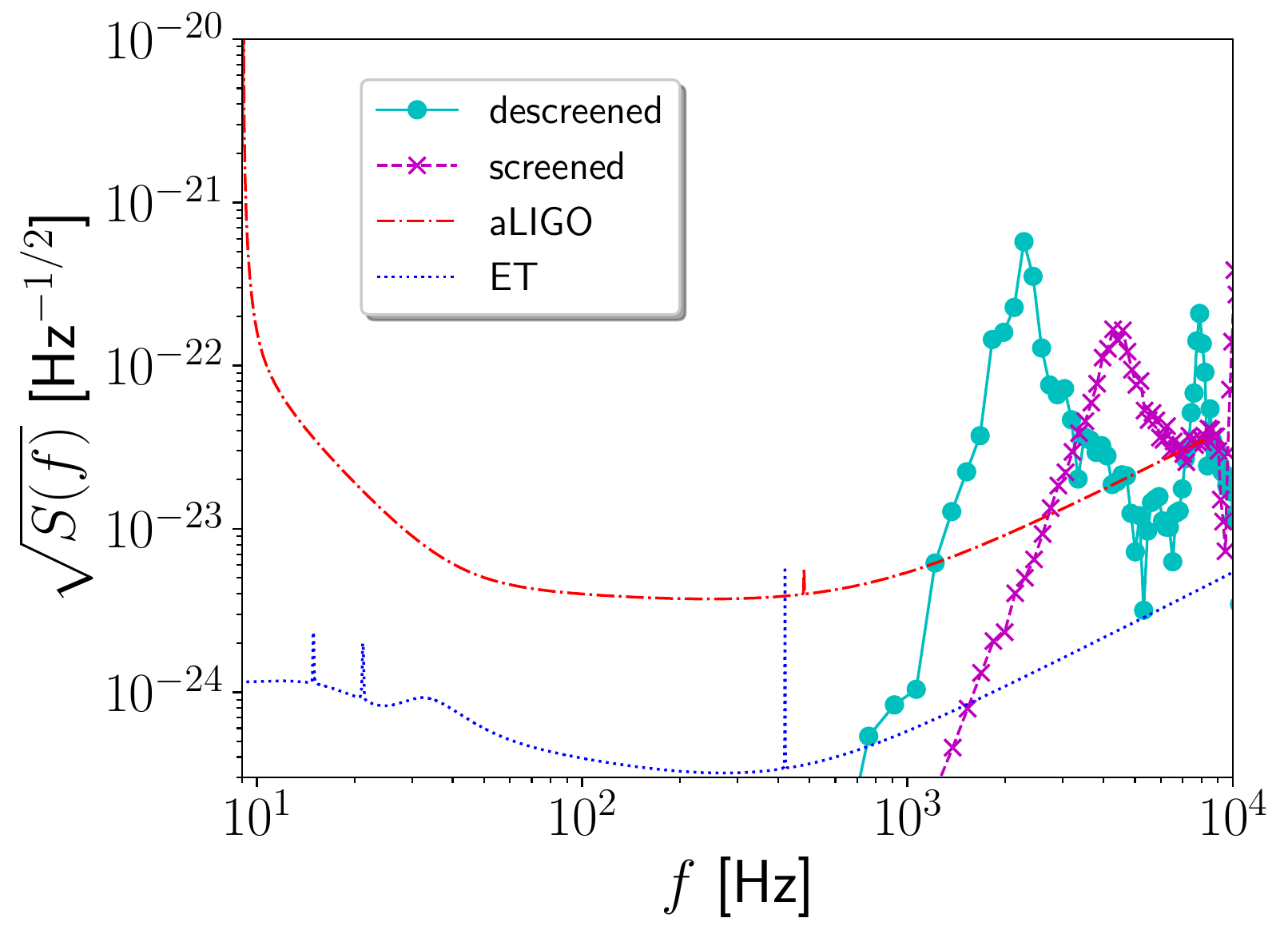}
\caption{ {\em{Signal vs detector sensitivity curves - oscillating stars.}} The strain amplitude in the frequency domain is compared to the design sensitivity curves of Advanced LIGO (red dot-dashed line) and ET (blue dotted line). The source-detector distance is set to $D_L=10$ kpc. The signals correspond to the monopole GWs produced by screened (continuous magenta line) and descreened (dashed cyan line) stars. The gravitational masses of the stars are, respectively, $M=1.02 M_{\odot}$ (screened CNS) and $M=1.84 M_{\odot}$ (descreened CNS). The initial perturbations (in the specific internal energy) employed to triggered the oscillations and scalar GWs emission have an amplitude of $\delta\xi_0=4\times 10^{-3}$. The visible peaks in the signals correspond to the fundamental mode, $F$, of the characteristic radial oscillations of the CNSs. The results have been obtained for $(\Lambda,\teinf)=(175$ GeV, $6.5\times 10^{10}$ g/cm$^3$).}
\label{fig:SNR}
\end{figure}

To investigate the observability of the GWs  sourced by the characteristic modes of matter inside oscillating CNSs (see Fig.~\ref{fig:qnms}), we compare the strain amplitude (in the frequency domain) of the signals produced by the screened and descreened stars
(both perturbed with the largest initial perturbation that we consider, $\delta\xi_0=4\times 10^{-3}$) with the sensitivity curves of Advanced LIGO and ET, as is shown in Fig.~\ref{fig:SNR}. We observe that only the fundamental mode $F$ (and, depending on the mass of the star, the first overtone $H_1$) have frequency falling in
the (high end of) the sensitivity range of ground-based detectors. We conclude that oscillating CNSs located within our Galaxy would produce signals that are well above the sensitivity threshold of Advanced LIGO, even in the case of the screened star, for the theory considered in these simulations. Conversely, oscillating stars located outside our Galaxy ($D_L\gtrsim$ Mpc) might be undetectable by Advanced LIGO (even in case of descreened CNSs) but within reach of third generation detectors such as the ET, for which we predict higher SNR values (see Table~\ref{tab:SNR}).
\begin{figure}[th!]
\centering
\includegraphics[width=0.5\textwidth]{ 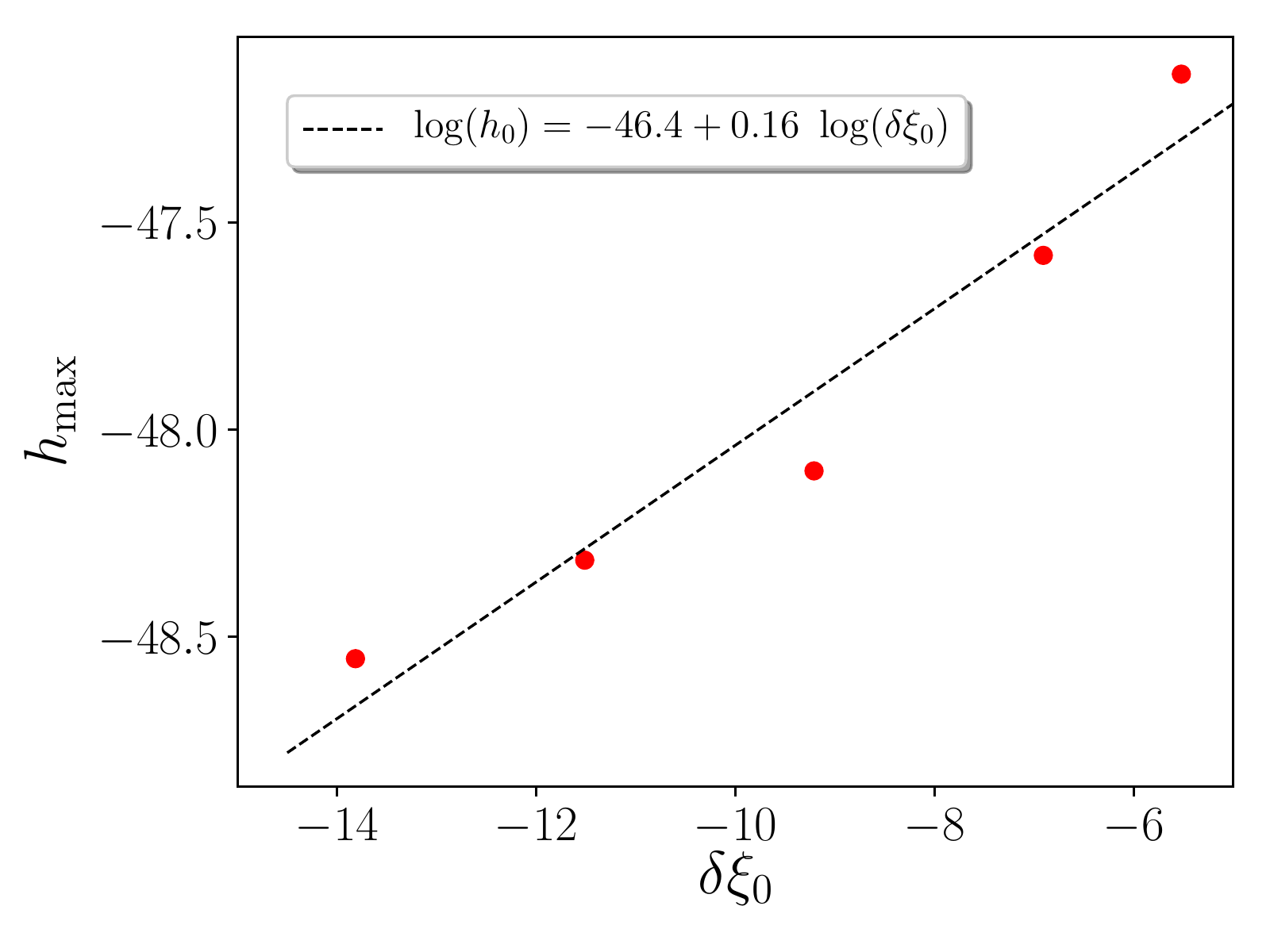}
\includegraphics[width=0.5\textwidth]{ 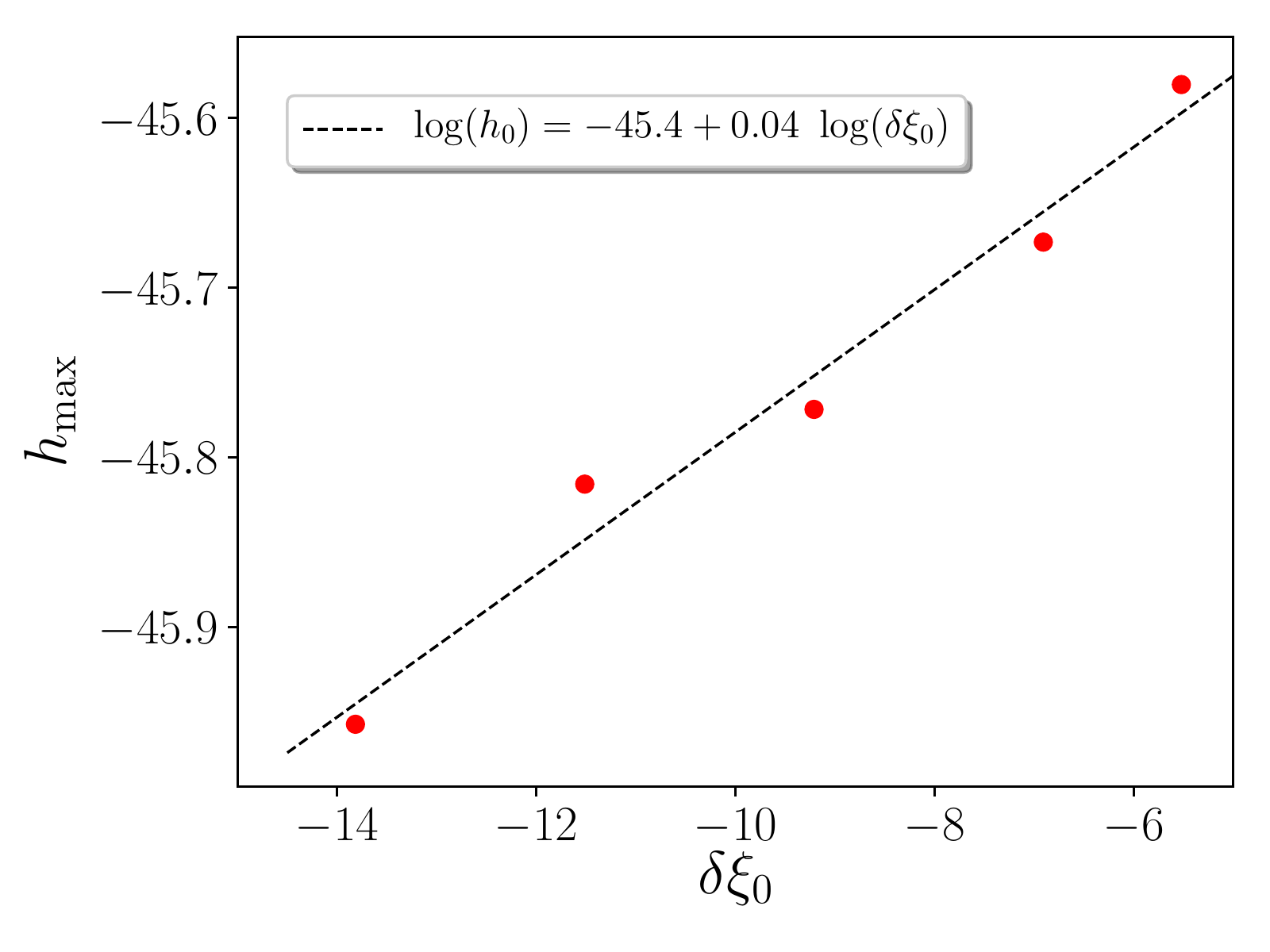}
\caption{ {\em{Scalar strain vs initial perturbation amplitude.}} The plots show the maximum amplitude of the monopole scalar radiation against the maximum amplitude of the initial perturbation of the specific internal energy, $\delta\xi$. Top and bottom panel correspond to a screened star with $M=1.02M_{\odot}$ and a descreened star with $M=1.84M_{\odot}$, respectively, located at a $D_L=10$ kpc distance from the detector. The amplitudes of the initial perturbation (in the specific internal energy) that we considered are $\delta\xi_0=\{10^{-6},10^{-5},10^{-4},10^{-3},4\times 10^{-3}\}$. The parameters of the model are set to $(\Lambda,\teinf)=(175$ GeV, $6.5\times 10^{10}$ g/cm$^3$). The black dashed lines show a power-law fit. }
\label{fig:h0vsamp}
\end{figure}

The scaling of our results with the the initial perturbation amplitude is shown (together with a power-law fit) in
Fig.~\ref{fig:h0vsamp}. As can be seen, the logarithmic dependence on the initial amplitude
suggests that our results are robust against changes in that quantity. We stress again, however,
that  all the results presented in this section have been obtained for $\Lambda\simeq175$ GeV.
When the chameleon energy scale is comparable to the dark-energy scale ($\sim$ meV), we expect the frequency of the $F$-mode to approach the GR predictions and thus to remain in the kHz range. However, the fundamental scalar mode, $F_s$, will have even higher frequencies because of the huge mass~\eqref{eq:mass} acquired by the chameleon field at nuclear densities, which may render detection of  scalar effects challenging.
As for the amplitude of the scalar signal, we expect it to be suppressed for $\Lambda\sim$ meV
and more realistic atmosphere densities. We will show this in detail for the (much stronger) scalar emission
produced in gravitational collapse, in the next section.

\begin{center}
\begin{table}[h!]
\begin{tabular}{|c|c|c|c|}
\hline
Scenario & Screening & LIGO & ET  \\
\hline\hline
&  &  &  \\[-1em]
\multirow{2}{*}{Oscillations} & Yes  & 4 &  $3.3\times 10^1$ \\
\cline{2-4}
&  &  &  \\[-1em]
& No & $2.0\times 10^1$ &  $1.6\times 10^2$\\
\hline
&  &  &  \\[-1em]
\multirow{2}{*}{Collapse} & Yes  & $7.6\times 10^3$ &  $7.8\times 10^4$ \\
\cline{2-4}
&  &  &  \\[-1em]
& No & $5.6\times 10^3$ &  $5.6\times 10^4$ \\
\hline
\end{tabular}
\caption{{\em{SNR.}} Estimates of the SNR of scalar GWs produced by oscillating and collapsing CNSs. Results are labelled by the presence or absence of screening in the core of the stars, and by the detector taken as a reference (Advanced LIGO or ET). The source-detector distance is set to $D_L=10$ kpc. The gravitational masses of the oscillating stars are, respectively, $M=1.02 M_{\odot}$ (screened CNS) and $M=1.84 M_{\odot}$ (descreened CNS). The collapsing CNSs have been chosen to have a fixed baryon mass $M_{bar}=1.75M_{\odot}$. The results are obtained for $(\Lambda,\teinf)=(175$ GeV, $6.5\times 10^{10}$ g/cm$^3$).}
\label{tab:SNR}
\end{table}
\end{center}

\subsection{Collapsing CNSs}
In this subsection, we extract the scalar (monopole) GW emission from simulations of collapsing unstable CNSs, respectively with and without descreened cores. In particular, we fixed the parameters of the theory to $(\Lambda,\teinf)=(175$ GeV, $6.5\times 10^{10}$ g/cm$^3$) and chose two CNSs with the same baryon mass (see Eq.~\eqref{eq:mbar}) $M_{bar}=1.75 M_{\odot}$; but with different EoS polytropic index, respectively $\Gamma=3$ and $\Gamma=2$.
For the latter value (and unlike for the former), the CNS does not feature a pressure-dominated core and the chameleon screening is fully effective. The collapse is induced with a small initial perturbation, introduced by decreasing the polytropic index by a tiny amount ($\sim 0.1\%$), which corresponds to a small increase of the initial pressure (by less than two percent) and of the specific internal energy (by half a percent).

\begin{figure}[t!]
\centering
\includegraphics[width=0.5\textwidth]{ 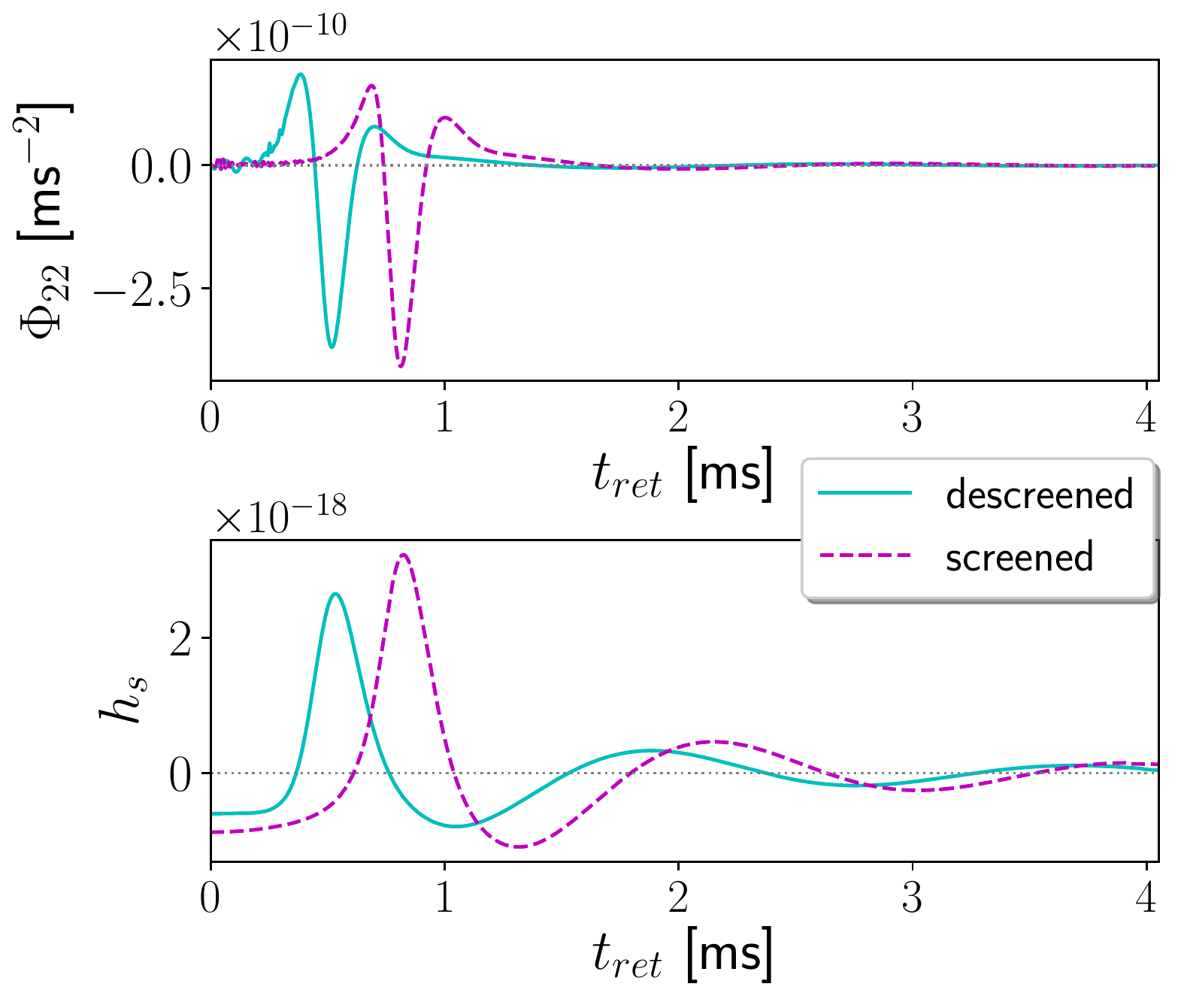}
\caption{ {\em{Scalar radiation from collapsing stars.}} Top panel: $\Phi_{22}$ vs retarded time. Bottom panel: strain amplitude vs retarded time. Bursts signals are extracted from simulations of collapsing CNSs, respectively with (dashed magenta lines) and without (continuous cyan lines) screening in the interior. The source-detector distance is set to $D_L=10$ kpc.}
\label{fig:gwcollapse}
\end{figure}

\begin{figure}[th!]
\centering
\includegraphics[width=0.5\textwidth]{ 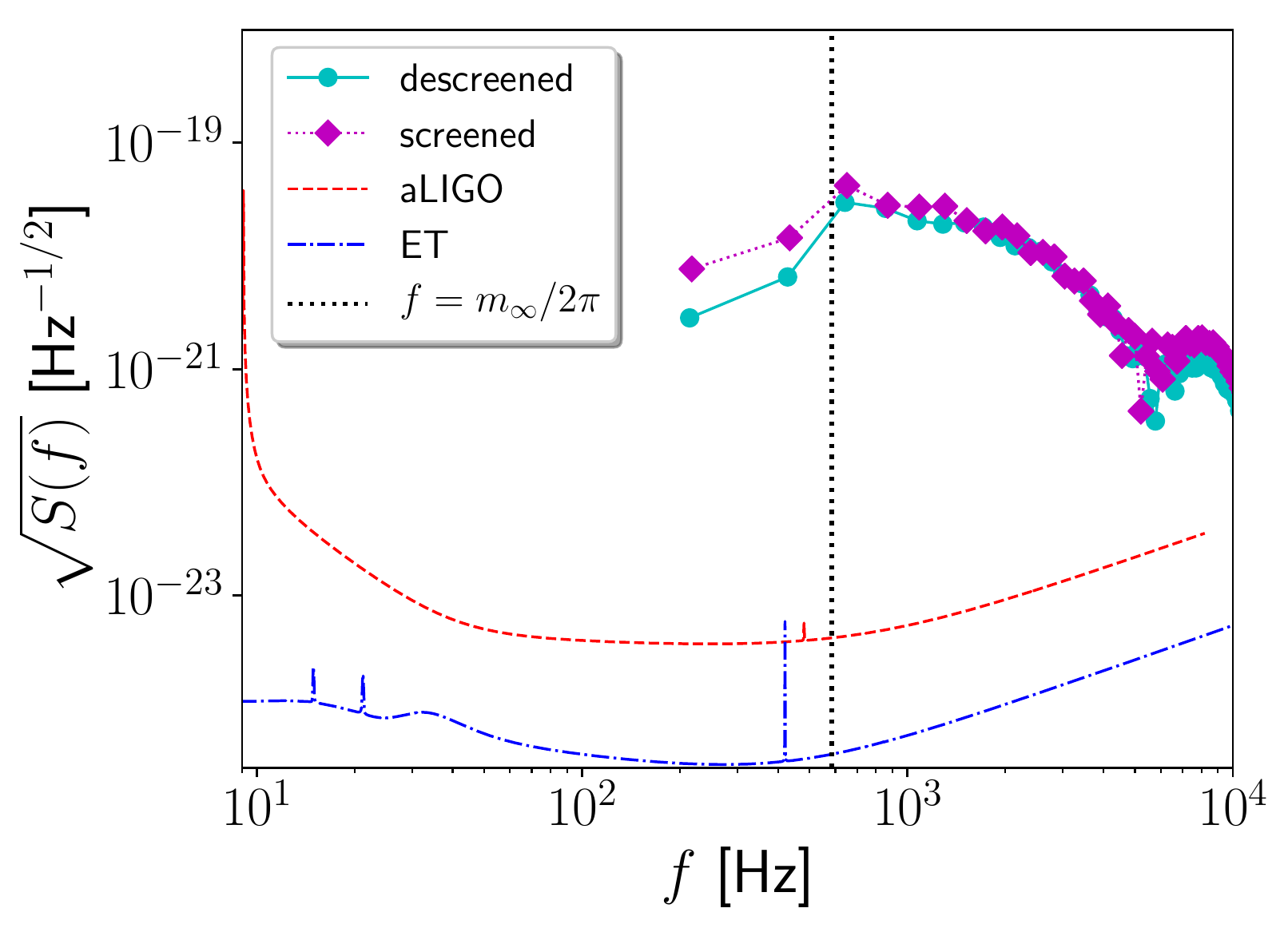}
\caption{ {\em{Signal vs detector sensitivity curves - collapsing stars.}} The strain amplitude in the frequency domain is compared to the design sensitivity curves of Advanced LIGO (red dot-dashed line) and ET (blue dotted line). The source-detector distance is set to $D_L=10$ kpc. The signals correspond to the monopole GWs produced by screened and descreened stars undergoing gravitational collapse. The vertical dash-dotted black line corresponds to $f_{\infty}=m_{\infty}/2\pi$, i.e. the peak frequency of the burst, below which all frequencies are Yukawa-suppressed.}
\label{fig:SNRcollapse}
\end{figure}

The plots in Fig.~\ref{fig:gwcollapse} show the monopole scalar GW produced by the two CNSs described above, at a distance of $D_L=10$ kpc.
The infalling matter produces a typical burst signal, visible in both the Newman-Penrose scalar $\Phi_{22}$ (top panel) and in the scalar strain amplitude $h_s$ (bottom panel). In these simulations we see no evidence of a suppression of the scalar emission due to screening (complete or partial).
In Fig.~\ref{fig:SNRcollapse} we compare the two scalar strain amplitudes, in the frequency domain, to the design sensitivity curves of current and next-generation terrestrial interferometers. As can be seen in the plot, a collapsing (screened or descreened) CNS would produce a very loud burst that would correspond to large SNR already in Advanced LIGO (see Table~\ref{tab:SNR}).


One may wonder, however, whether this large monopole radiation persists for smaller values of  $(\Lambda,\teinf)$.
To answer this question, let us try to gain some insight on why
large scalar signals are produced in our simulations.
As mentioned in Sec.~\ref{three}, the end state of the collapse of a CNS is a ``hairless'' BH with the chameleon field lying in the constant ``exterior''
vacuum, $\phi=\phi_{\infty}$. Note indeed that vacuum solutions with ``hair'' (i.e. non-constant scalar field) are forbidden by a trivial generalization of the Hawking-Bekenstein ``no-scalar-hair'' theorem~\cite{Hawking:1972qk,Bekenstein:1972ny,Bekenstein:1995un}. As a result, the scalar charge of the star must be shed away via GW emission during collapse. Therefore, larger initial charges will correspond to larger burst amplitudes. Note that a similar mechanism, whereby gravitational collapse has to shed away (because of no-hair theorems) any scalar hair that a star may initially have, thus producing a strong scalar monopole emission, was recently discovered for theories that yield kinetic screening~\cite{Bezares:2021yek}.

In our case, we observe that at large values of $(\Lambda,\teinf)$ the scalar charges of CNSs are not efficiently suppressed by the ``thin-shell'' effect. In fact, one can notice that the screening radius (see sec.~\ref{twoB} for the definition) of the TOV solutions obtained by choosing $(\Lambda,\teinf)=(175$ GeV$, 6.5\times 10^{10}$ g/cm$^3)$ is typically $\lesssim 70\%$ of the size of the stars (see Fig.~\ref{fig:descreen} and also Fig.~$2$ in~\cite{deAguiar:2020urb}). The relativistic stars are thus in a ``thick-shell'' regime, i.e. a non-negligible fraction of the stellar mass sources the scalar charge. In our simulations, in particular, the screened and descreened CNSs shown in Fig.~\ref{fig:gwcollapse} emit loud scalar GWs because they  have relatively large and comparable charges, respectively $Q\simeq0.15$ and $Q\simeq0.11$. The descreened star actually features a charge slightly smaller than the screened CNS. We interpret this as due to the descreened core, which gives a negative contribution to charge and thus decreases the its total value.

\begin{figure}[t!]
\centering
\includegraphics[width=0.5\textwidth]{ 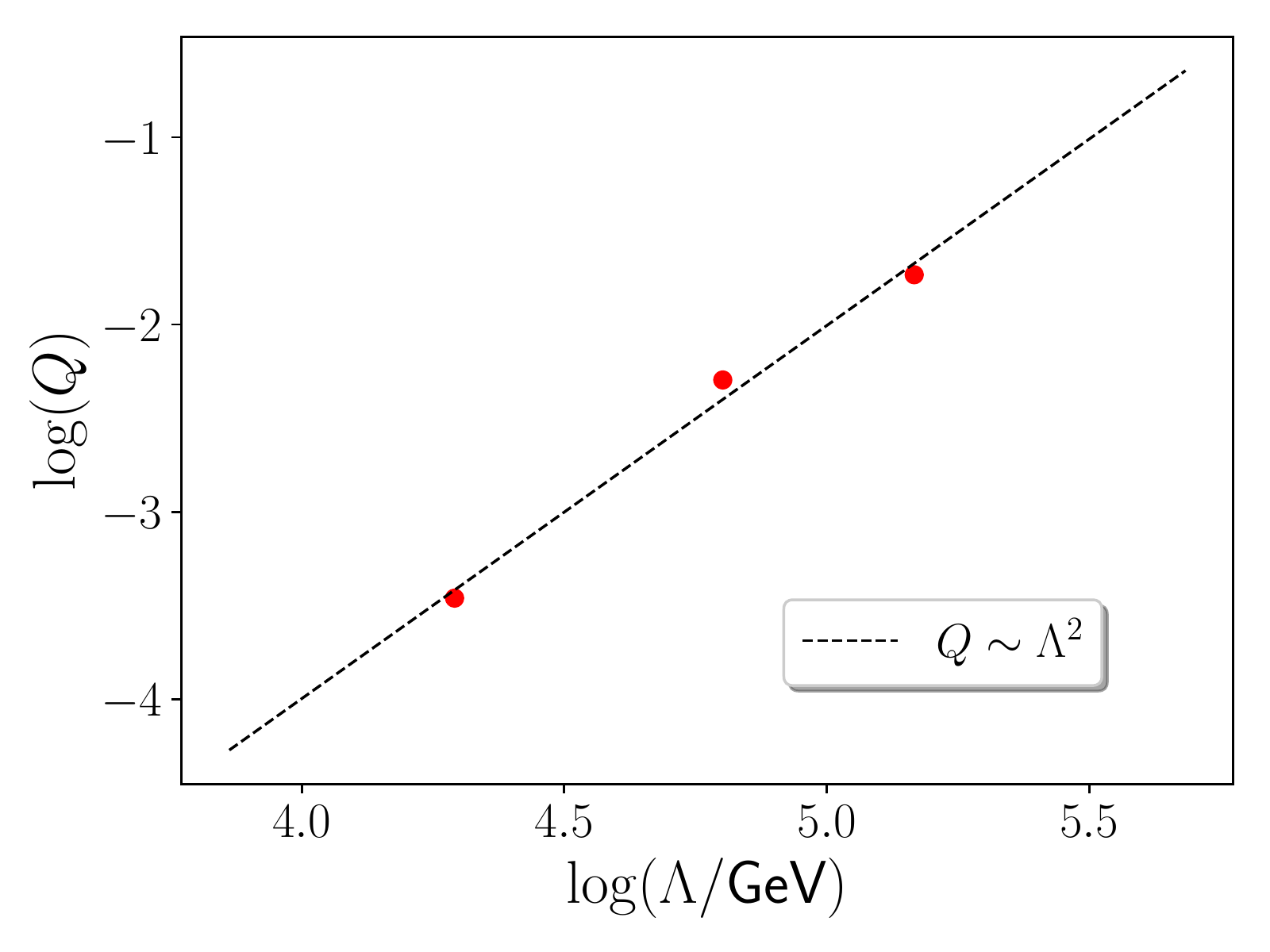}
\caption{ {\em{Scalar charge vs chameleon energy scale}}. The plot shows the scalar charge, $Q$, of CNSs with $M_{bar}=1.75M_{\odot}$ against chameleon energy scale, $\Lambda$; the atmosphere density is kept constant, $\teinf=6.5\times 10^{10}$ g/cm$^3$. Red dots represent data corresponding to $\Lambda=\{175,122,73\}$ GeV. The black dashed line represents the power law $Q\sim\Lambda^{a}$ (with $a\simeq 2$) fitting the data.}
\label{fig:QLam}
\end{figure}

\begin{figure}[t!]
\centering
\includegraphics[width=0.5\textwidth]{ 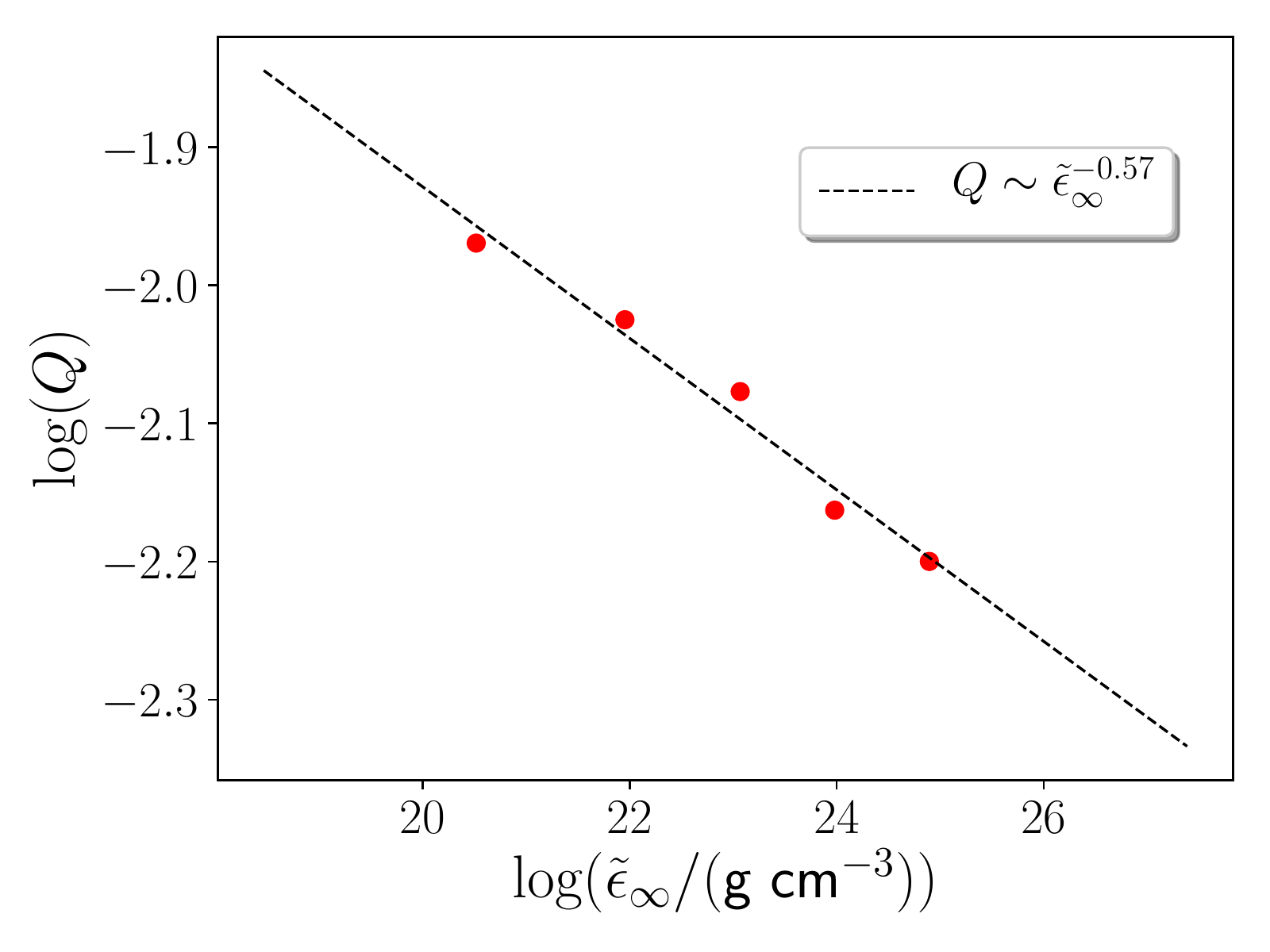}
\caption{ {\em{Scalar charge vs atmosphere density}}. The plot shows the scalar charge, $Q$, of CNSs with $M_{bar}=1.75 M_{\odot}$ against atmosphere density, $\teinf$; the asymptotic value of the chameleon field is fixed to $\phi_{\infty}\simeq 0.17 M_{\textrm{pl}}$. Red dots represent data corresponding to $\teinf=\{6.5,2.6,1.1,0.34,0.081\}\times 10^{10}$ g/cm$^3$. The black dashed line represents the power-law $Q\sim\teinf^b$ (with $b\simeq-3/5$) fitting the data.}
\label{fig:QAtm}
\end{figure}

To extrapolate the charges of CNSs to realistic values of $(\Lambda,\teinf)$, we use the scaling
\begin{equation}\label{eq:Qscaling}
Q(\Lambda,\teinf)=(\Lambda/\Lambda_0)^{a}(\teinf/\teps_0)^{b}Q(\Lambda_0,\teps_0)\,,
\end{equation}
where the coefficients $a\simeq 2$ and $b\simeq -3/5$ were obtained by power-law fits
of simulations with baryon mass $M_{bar}=1.75M_{\odot}$ (c.f. Figs~\ref{fig:QLam}--\ref{fig:QAtm}).
Based on Eq.~\eqref{eq:Qscaling}, we predict that the CNSs of mass $M_{bar}=1.75M_{\odot}$ considered above will have a scalar charge of, respectively, $Q\simeq6\times 10^{-11}$ and $Q\simeq5\times 10^{-11}$ for the realistic values  $(\Lambda,\teinf)\simeq (2.4$ meV$,1.67\times 10^{-20}$ g/cm$^{3})$. We interpret this suppression of the scalar charge as a vindication of the ``thin-shell'' effect, which appears to be restored, even for relativistic stars, at realistic values of the parameters of the theory.

\begin{figure}[th!]
 \centering
 \includegraphics[width=0.5\textwidth]{ 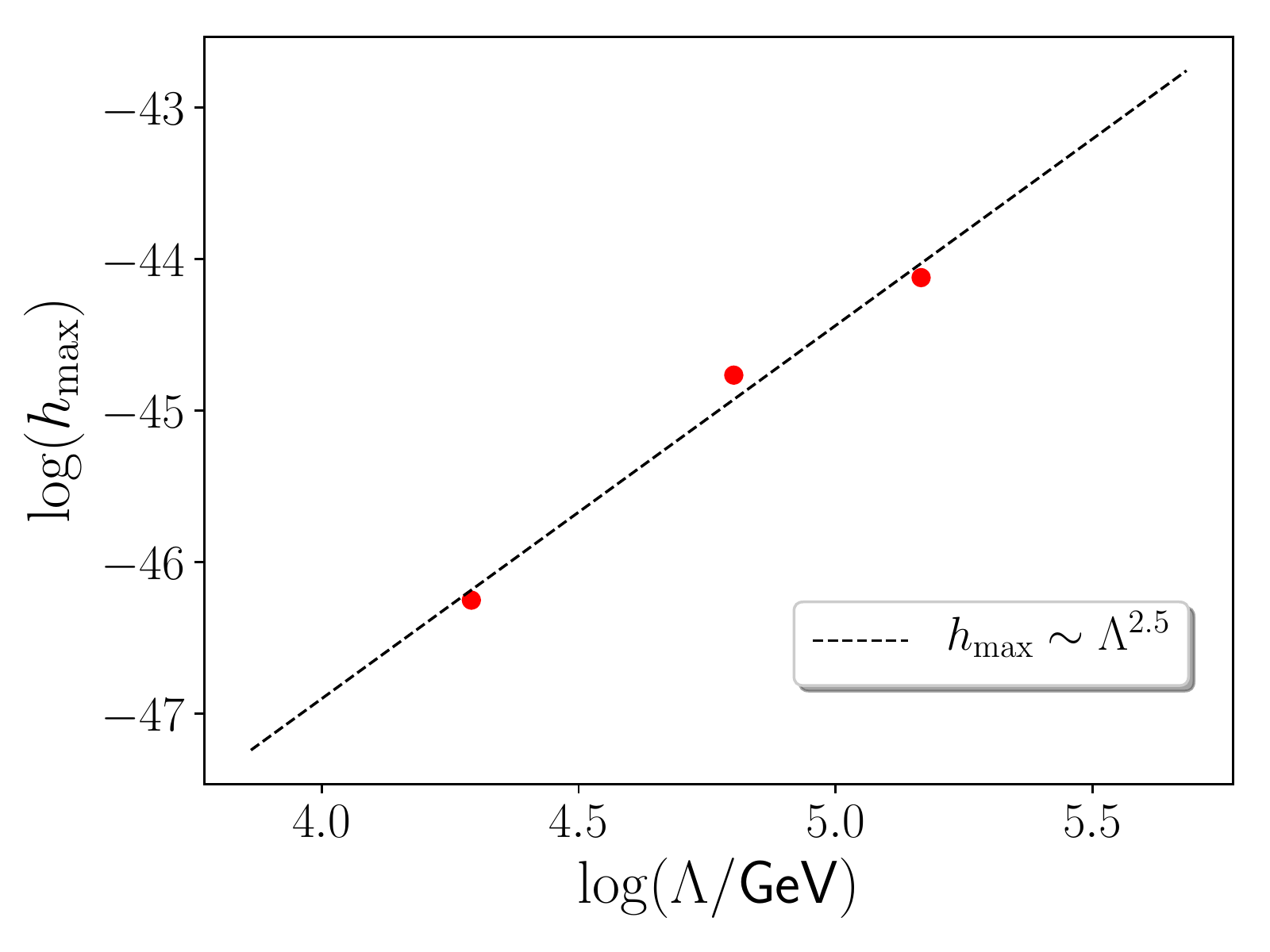}
 \caption{ {\em{Maximum burst amplitude vs chameleon energy scale}}. The plot shows the maximum amplitude, $h_{\max}$, of the burst produced by collapsing CNSs with $M_{bar}=1.75 M_{\odot}$ against the chameleon energy scale, $\Lambda$; the atmosphere density is kept constant, $\teinf=6.5\times 10^{10}$ g/cm$^3$. Red dots represent data corresponding to $\Lambda=\{175,122,73\}$ GeV. The black dashed line represents the power law $h_{\max}\sim\Lambda^{c}$ (with $c\simeq 5/2$) fitting the data.}
 \label{fig:h0Lam}
 \end{figure}

\begin{figure}[th!]
 \centering
 \includegraphics[width=0.5\textwidth]{ 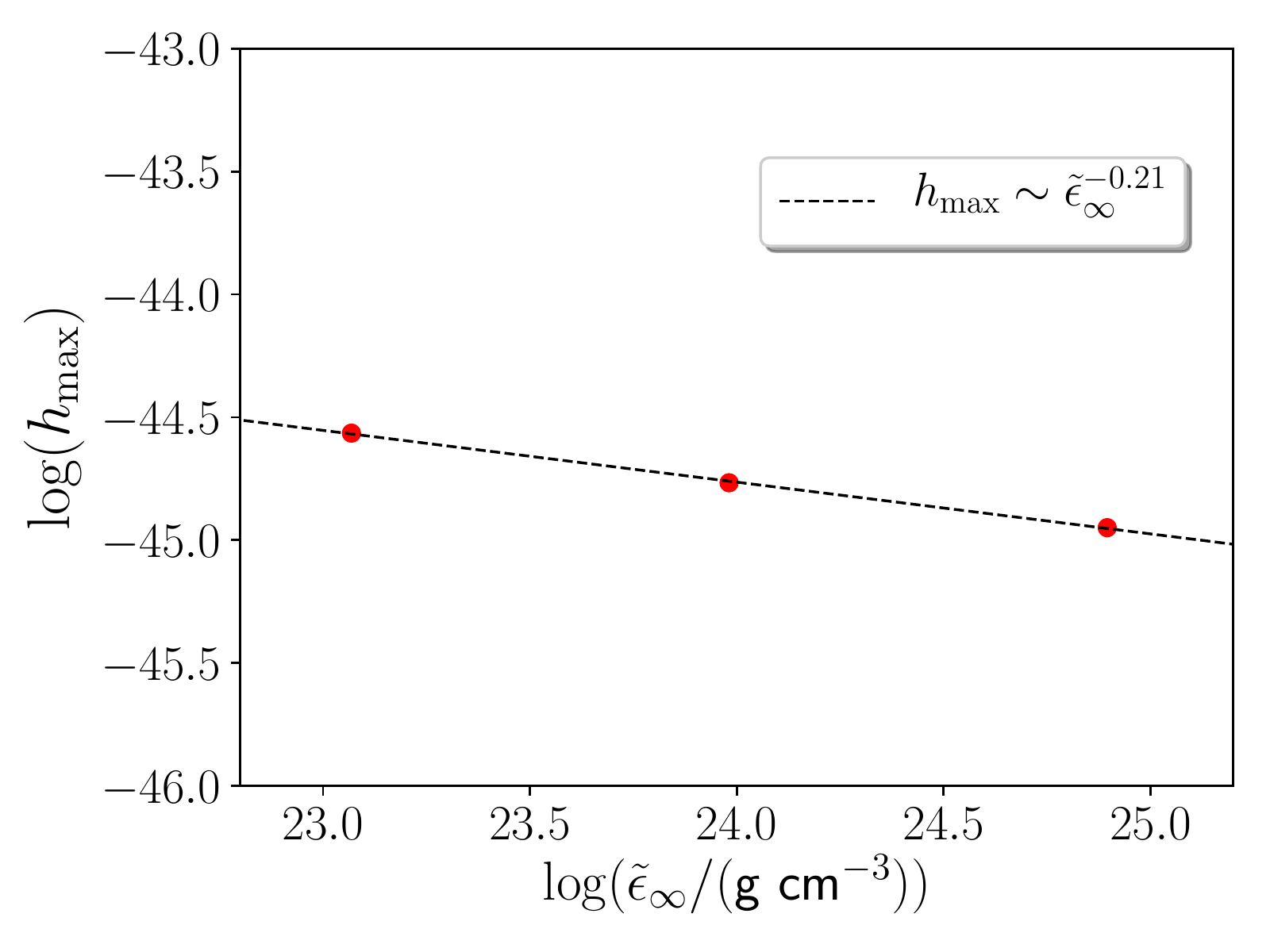}
 \caption{ {\em{Maximum burst amplitude vs atmosphere density}}. The plot shows the maximum amplitude, $h_{\max}$, of the scalar GW burst against the atmosphere density, $\teinf$. The collapsing CNSs have a fixed baryon mass $M_{bar}=1.75 M_{\odot}$ and the asymptotic value of the chameleon field is fixed to $\phi_{\infty}\simeq 0.17 M_{\textrm{pl}}$. Red dots represent data corresponding to $\teinf=\{6.5,2.6,1.1\}\times 10^{10}$ g/cm$^3$. The black dashed line represents the power law $h_{\max}\sim\teinf^d$ (with $d\simeq-7/10$) fitting the data.}
 \label{fig:h0Atm}
 \end{figure}

Motivated by this result, we turn now to estimating the SNR of burst signals for realistic/viable values of $(\Lambda,\teinf)$.
To overcome the technical challenges of directly simulating stars at very small $\Lambda$ and $\teinf$ (see discussion in Sec.~\ref{four}),
we resort again to determining the scaling of the scalar monopole signal with these quantities. From simulations of the collapse with $\Lambda\simeq\{175, 122, 73\}$ GeV and $\teinf=6.5\times 10^{10}$ g/cm$^3$,
we fit the maximum strain amplitude of the burst as a function of $\Lambda$ using a power law, as displayed in Fig.~\ref{fig:h0Lam}.
We then fit (again with a power law) the same quantity against the exterior density, $\teinf=\{6.5,2.6,1.1\}\times 10^{10}$ g/cm$^3$, using simulations with fixed $\phi_{\infty}=0.17M_{\textrm{pl}}$. The result is shown in Fig.~\ref{fig:h0Atm}.

 Combining the results from these power-law fits, one obtains a scaling relation for the maximum scalar amplitude
\begin{align}
h_s(\Lambda,\teinf,D_L;&t_{ret})\simeq
(\Lambda/\Lambda_0)^{c}(\teinf/\teps_0)^{d}(D_0/D_L)~\cdot\nonumber\\
&\cdot~h_s(\Lambda_0,\teps_0,D_0;(m_0/m_{\infty}) \,t_{ret})\,,
\label{eq:hscaling}
\end{align}
with $c\simeq5/2$ and $d\simeq-7/10$.
Note that the scaling with $\Lambda$ coincides with that of the quantity $\phi_{\infty}-\phi_s$,
where $\phi_s$ is the minimum of the scalar field inside the CNS. Indeed, from Eq.~\eqref{eq:phimin} one obtains $\phi_{\infty}$, $\phi_s\sim \Lambda^{5/2}$.
Let us also note, as can be seen from Fig.~\ref{fig:SNRcollapse}, that the burst signal peaks at $f=f_{\infty}\equiv m_{\infty}(\Lambda,\teinf)/2\pi$, while lower frequencies are suppressed. Making use of expression~\eqref{eq:mass}, one can check that lower values of $\Lambda$ and $\teinf$ correspond to smaller chameleon masses, and thus lower peak frequencies. Hence, to extrapolate to lower $(\Lambda,\teinf)$ one also needs to rescale the time by the factor $(m_0/m_{\infty})$ that appears in Eq.~\eqref{eq:hscaling}.

Finally, by applying Eq.~\eqref{eq:hscaling} to extrapolate to $(\Lambda,\teinf)\simeq (2.4$ meV$,1.67\times 10^{-20}$ g/cm$^{3})$~
(the latter corresponding to the order of magnitude of the density inside large molecular clouds),
we find that the monopole signal would peak in the mHz band, outside the band of terrestrial detectors
but within that of LISA. Although we have computed the SNR for LISA,  that is completely
unobservable ($\textrm{SNR}\simeq10^{-10}$), even for distances of a few kpc.

\subsection{Binary systems}
From the extrapolation presented in the previous section, we have concluded
that the monopole emission from collapsing CNSs is pratically unobservable with current (and future) detectors, at least for realistic
values of the chameleon model.
When it comes to (quasi-circular) binary systems involving at least one NS, the strongest effect
is expected to be dipole scalar emission, which potentially dominates the binary's evolution at low frequencies~\cite{Damour:1992we,Damour:1996ke,Barausse:2012da,Freire:2012mg,Palenzuela:2013hsa,Wex:2014nva,Barausse:2016eii,Sagunski:2017nzb},
The deviations from GR induced by dipole emission can be parametrized via~\cite{Barausse:2016eii}
\begin{equation}\label{eq:Eloss}
\dot{E}=\dot{E}_{GR}\left(1+\frac{B}{v^2}\right)\,,
\end{equation}
where $v$ is the relative velocity of the binary,
$\dot{E}$ and $\dot{E}_{GR}$ are the total energy fluxes in chameleon gravity and in GR, respectively,
and $B\sim(Q_1-Q_2)^2$ (with $Q_1$ ad $Q_2$ the component charges).

Note that Eq.~\eqref{eq:Eloss} is valid for ST theories with a massless scalar, while
the chameleon field possesses a non-vanishing mass. We therefore
expect the energy loss due to dipole radiation to be given by Eq.~\eqref{eq:Eloss}  only at
binary separations smaller than the Compton wavelength. For $(\Lambda,\teinf)\simeq (2.4$ meV$,1.67\times 10^{-20}$ g/cm$^{3})$,
the Compton wavelength is $\lambda_c=1/m_{\infty}\simeq O(10^8)$ km, which is larger than the typical separation of binary pulsars
(which is $\lesssim 10^6$ km). However, from the scaling~\eqref{eq:Qscaling}, the scalar charge of relativistic stars extrapolated at
$(\Lambda,\teinf)\simeq (2.4$ meV$,1.67\times 10^{-20}$ g/cm$^{3})$ would be $Q\lesssim O(10^{-10})$, corresponding to
$B\sim 10^{-20}$. This is at least 10 orders of magnitude lower than what is currently measurable~\cite{Barausse:2016eii}

\section{Conclusion}\label{conclusion}
In this paper, we have investigated the chameleon screening mechanism in the fully dynamical and nonlinear regime
of oscillating and collapsing NSs, in spherical symmetry. Our simulations confirm the static results of Ref.~\cite{deAguiar:2020urb}, and in particular
the partial breakdown of the chameleon screening inside stars with pressure-dominated cores,
but also provide evidence of the nonlinear stability of both screened and partially descreened stars in chameleon gravity.

We have focused first on the characteristic spectrum of (radial) oscillations of NSs. We observed a shift in the frequencies of the fundamental mode and higher overtones with respect to the GR predictions. While this effect could be degenerate with the EoS, the appearance of a new family of modes
may potentially constitute the ``smoking gun'' of a gravitational scalar field.
However, these modes have frequencies of the order of the large mass that the chameleon field acquires inside relativistic stars (i.e. $\gtrsim$ kHz).
Moreover, we find that chameleon screening is also quite efficient at suppressing
the scalar mode amplitude in oscillating screened CNSs, already at large $\Lambda\sim 100$ GeV.  For this reason,
scalar effects in oscillating stars are probably unobservable for realistic chameleon energy scales $\Lambda\simeq$ meV.

We have also simulated gravitational collapse of NSs,
which can lead to larger monopole scalar signals than stellar oscillations. We have assessed detectability
by existing and future GW interferometers, concluding that
the scalar radiation would be observable in the Galaxy for large chameleon energy scales $\Lambda \sim 100$ GeV.
However, if one extrapolates our results down to viable chameleon energy scales $\Lambda\simeq$ meV,
the screening suppresses the amplitude of the signal, which also gets shifted to lower
($\sim$ mHz) frequencies. We have checked that, as a result, this scalar emission would be undetectable
even with LISA. Similarly, our results for the scalar charge of isolated NSs suggest that
scalar effects would be suppressed by screening also in pulsar binary systems for $\Lambda\simeq$ meV.

\begin{acknowledgments}
We are grateful to  C. Palenzuela  for illuminating discussions about
spherical collapse and numerical relativity, and to  Raissa Mendes
for insightful conversations on chameleon stars and for reading a preliminary version of this manuscript.
We acknowledge financial support provided under the European Union's H2020 ERC Consolidator Grant
``GRavity from Astrophysical to Microscopic Scales'' grant agreement no. GRAMS-815673.
\end{acknowledgments}

\bibliographystyle{utphys}
\bibliography{biblio}

\end{document}